\documentclass[preprint,prd,nofootinbib,tightenlines,amsmath]{revtex4}
\usepackage{axodraw}
\usepackage{graphics}
\usepackage{epsfig}
\usepackage{dcolumn}
\usepackage{bm}

\begin{document}
\baselineskip=15pt \parskip=5pt

\hspace*{\fill} $\hphantom{-}$

\def\lsim{\mathrel {\vcenter {\baselineskip 0pt \kern 0pt
    \hbox{$<$} \kern 0pt \hbox{$\sim$} }}}
\def\gsim{\mathrel {\vcenter {\baselineskip 0pt \kern 0pt
    \hbox{$>$} \kern 0pt \hbox{$\sim$} }}}

\title{Light Higgs Production in Hyperon Decay}

\author{Xiao-Gang He}
\email{hexg@phys.ntu.edu.tw} \affiliation{Department of Physics
and Center for Theoretical Sciences, National Taiwan University,
Taipei.}

\author{Jusak Tandean}
\email{jtandean@ulv.edu}
\affiliation{Department of Mathematics/Physics/Computer Science,
University of La Verne, La Verne, CA 91750, USA}

\author{G. Valencia}
\email{valencia@iastate.edu}
\affiliation{Department of Physics and Astronomy, Iowa State University, Ames, IA 50011, USA}

\date{\today}

\begin{abstract}

A recent HyperCP observation of three events in the decay
$\Sigma^+\to p\mu^+\mu^-$  is suggestive of a new particle with mass 214.3\,MeV.
In order to confront  models that contain a light Higgs boson with this observation,
it is necessary to know the Higgs production rate in hyperon decay.
The contribution to this rate from penguin-like two-quark operators has been considered
before and found to be too large. We point out that there are additional four-quark
contributions to this rate that could be comparable in size to the two-quark contributions,
and that could bring the total rate to the observed level in some models.
To this effect we implement the low-energy theorems that dictate the couplings of light
Higgs bosons to hyperons at leading order in chiral perturbation theory.
We consider the cases of scalar and pseudoscalar Higgs bosons in the standard model
and in its two-Higgs-doublet extensions to illustrate the challenges posed by existing
experimental constraints and suggest possible avenues for models to satisfy them.
\end{abstract}


\maketitle

\section{Introduction}

Three events for the decay mode \,$\Sigma^+\to p\mu^+\mu^-$\,  with a dimuon invariant mass of
\,$214.3\pm0.5$\,MeV\, have been recently observed by the HyperCP Collaboration~\cite{Park:2005ek}.
It is possible to account for these events within the standard model (SM) when long-distance
contributions are properly included~\cite{He:2005yn}.
However, the probability of having all three events at the same dimuon mass in the SM
is less than one percent.
This suggests a new-particle interpretation for the events, for which the branching ratio
is  \,$\bigl(3.1^{+2.4}_{-1.9}\pm1.5\bigr)\times 10^{-8}$~\cite{Park:2005ek}.\,

This possibility has been explored to some extent in the literature, where it has been
shown that kaon decays place severe constraints on the couplings of the hypothetical
new particle~\cite{He:2005we,Deshpande:2005mb,Geng:2005ra}.
In particular, it was found that the flavor-changing coupling of the new state, $X$, to
$\bar{d}s$  has to be of a pseudoscalar or axial-vector nature to explain why the state
has not been seen in  \,$K\to\pi\mu^+\mu^-$.\,
At least one model containing a particle with these properties has
appeared in the literature~\cite{Gorbunov:2000cz}.

All these previous analyses of $X$ considered only the effects of two-quark operators
for  $\bar d sX$.
However, it is well known in the case of light-Higgs production in kaon decay that there
are also four-quark operators that can contribute at the same level as the two-quark
ones~\cite{sdH,Leutwyler:1989xj,Gunion:1989we,Grzadkowski:1992av}.
These four-quark contributions are most conveniently described in chiral perturbation
theory ($\chi$PT) which implements low-energy theorems governing the couplings of light
(pseudo)scalars to hadrons.
In this paper we generalize existing studies appropriate for kaon decay to the case of
hyperon decay.
This allows us to discuss the production of light (pseudo)scalars in hyperon decay
consistently, including the effects of both the two- and four-quark operators  with
the aid of~$\chi$PT.
We consider the cases of scalar and pseudoscalar Higgs bosons in the SM and in the
two-Higgs-doublet model~(2HDM), expressing our results in a form that can be
easily applied to more complicated Higgs models.

This paper is organized as follows. We begin  by collecting in Sec.~\ref{constraints}
the existing constraints on light Higgs bosons from kaon, $B$-meson, and
hyperon decays if we interpret the HyperCP events as being mediated by a light Higgs boson.
In Secs.~\ref{scalarH} and~\ref{pseudoscalarH} we compute the production rates in both
kaon and hyperon decays for a light scalar and pseudoscalar Higgs boson, respectively.
Finally in Sec.~\ref{final} we summarize our results and state our conclusions.

\section{Summary of Existing Constraints\label{constraints}}

In Ref.~\cite{He:2005we} we parameterized the possible couplings of the new particle, $X$,
to $\bar d s$ and $\bar\mu\mu$ assuming that it had definite parity.
Whereas this is a reasonable assumption for the diagonal couplings of $X$ to fermions,
it is not for its flavor-changing neutral couplings (FCNCs).
Two-quark FCNCs are predominantly induced by Higgs-penguin diagrams, which result in left- and
right-handed couplings, implying that the scalar and pseudoscalar ones are present simultaneously.
For this reason, we revisit the existing constraints for $X$ being a scalar particle, ${\cal H}$,
or a pseudoscalar particle, ${\cal A}$, assuming them to have two-fermion FCNCs described by
\begin{subequations}  \label{quark}
\begin{eqnarray}  \label{Hsd}
{\cal L}_{{\cal H}sd}^{} &=& \frac{g_{\cal H}^{}}{v} \left[m_s^{}\,\bar d(1+\gamma_5^{})s
\,+\,  m_d^{}\,\bar{d}(1-\gamma_5^{})s \right]{\cal H} \,\,+\,\, {\rm H.c.}  \,\,,
\\ \label{Asd}
{\cal L}_{{\cal A}sd}^{} &=&
\frac{ig_{\cal A}^{\vphantom{\sum}}}{v} \left[m_s^{}\,\bar d(1+\gamma_5^{})s
\,-\,  m_d^{}\,\bar{d}(1-\gamma_5)s \right]{\cal A} \,\,+\,\, {\rm H.c.}  \,\,,
\end{eqnarray}
\end{subequations}
where  the $g$'s are coupling constants,  $m_q^{}$  is a quark mass, and
\,$v=2^{-1/4}\,G_F^{-1/2}= 246\,{\rm GeV}$.\,
In addition, the diagonal couplings to charged leptons are assumed to have definite
parity and be proportional to the lepton mass,
\begin{eqnarray}
{\cal L}_{{\cal H}\ell}^{} \,\,=\,\,
\frac{g_\ell^{}\, m_\ell^{}}{v}\,\bar\ell \ell\, {\cal H}  \,\,,  \hspace{2em}
{\cal L}_{{\cal A}\ell}^{} \,\,=\,\,
\frac{i g_\ell^{}\, m_\ell^{}}{v}\,\bar\ell \gamma_5 \ell\, {\cal A}  \,\,.
\end{eqnarray}

For a (pseudo)scalar of mass 214.3\,MeV, it is then natural to assume that the decay
\,$X\to\mu^+\mu^-$\,  will dominate over the other kinematically allowed modes:
\,$X\to e^+e^-,\,\nu\bar\nu,\,\gamma\gamma$.\,
We will restrict ourselves to this case, assuming that  \,${\cal B}(X\to\mu^+\mu^-)\sim 1$.\,
This is true, for example, for a~light SM Higgs boson where  \,$g_\ell^{}=1$,\, or for light
pseudoscalars in the 2HDM types I and II, where  \,$g_\ell^{}=\cot\beta$ and $-\tan\beta$,
respectively.
In all these cases decays,  \,$X\to e^+e^-$\,  are suppressed at least by
\,$(m_e^{}/m_\mu^{})^2\sim 10^{-5}$.

To be consistent with the HyperCP observation, $X$ must be short-lived and
decay inside the detector.
This is compatible with the estimate for the total width
\,$\Gamma_{\cal A}\sim 10^{-7}$\,MeV~\cite{Geng:2005ra} of a pseudoscalar particle, ${\cal A}$.
It was shown in Ref.~\cite{He:2005we} that the muon anomalous magnetic moment imposes
the constraint
\begin{equation}
|g_\ell^{}| \,\,\lesssim\,\, 1.2  \,\,.\label{gmu}
\end{equation}
A coupling satisfying this constraint implies a width
\begin{equation}
\Gamma_{\cal A}^{}  \,\,\lesssim\,\,  3.7 \times 10^{-7}~{\rm MeV}\,\,,
\end{equation}
consistent with the observation. In contrast, the corresponding constraint for a scalar
particle is  \,$|g_\ell^{}| \lesssim 0.98$,\,  leading to a longer lifetime,
\begin{equation}
\Gamma_{\cal H}^{}  \,\,\lesssim\,\,  6.9 \times 10^{-9}~{\rm MeV}\,\,.
\end{equation}
The estimated lifetime for the HyperCP particle is therefore consistent with that of
a pseudoscalar or scalar that decays predominantly into muons.

In addition to the two-quark contributions to the amplitudes for
\,$K\to\pi{\cal H(A)}$\,  and  \,$\Sigma^+\to p{\cal H(A)}$\,  induced by
the interactions in Eq.~(\ref{quark}), we will also include contributions arising from the usual
SM four-quark  \,$|\Delta S|=1$\, operators, along with flavor-conserving couplings of
$\cal H(A)$.
We will adopt the chiral-Lagrangian approach to evaluate the hadron-level interactions.

Later on we will discuss specific models and consider the bounds
appropriate for them, including all the relevant two- and four-quark contributions.
It is useful to start with one example to illustrate the ingredients needed to construct
a model that can satisfy all the existing constraints.
For this purpose, we consider a pseudoscalar ${\cal A}$  with two-quark couplings as in
Eq.~(\ref{Asd}) supplemented with simple parameterizations for the
four-quark amplitudes for both kaon and hyperon decays.
For $B$-meson decay, we assume that the two-quark contribution completely dominates.

\subsection{$\bm{K\to\pi{\cal A}}$}

Introducing the dimensionless quantity $M_{4K}$ for the four-quark contribution, we express
the amplitude for  \,$K^\pm\to\pi^\pm\cal A$\,  and its branching ratio, respectively, as
\begin{eqnarray}
i{\cal M}(K^\pm\to\pi^\pm{\cal A}) &=&
g_{\cal A}^{}\,\frac{m_K^2-m_\pi^2}{v} \,-\, M_{4K}^{}\,\frac{m_K^2}{v} \,\,,
\nonumber \\
{\cal B}(K^\pm\to\pi^\pm{\cal A}) &=&
4.43\times10^{8} \left|g_{\cal A}^{}-1.08\,M_{4K}^{}\right|^2  \,\,.
\label{keq}
\end{eqnarray}
This mode is constrained by its nonobservation in the BNL E865~\cite{Ma:1999uj} or
FNAL HyperCP~\cite{Park:2001cv} measurements of  \,$K^\pm\to\pi^\pm\mu^+\mu^-$.\,
It is also constrained by its nonobservation  in the isospin-related mode
\,$K_S\to\pi^0\mu^+\mu^-$\,  by CERN NA48~\cite{Batley:2004wg}.
Of these three experiments, E865 had the best statistics, collecting 430 events in
\,$K^+\to\pi^+\mu^+\mu^-$.\,  A new particle ${\cal A}$  of mass  214.3\,MeV would have
contributed only in their first dimuon-mass bin, where
\,$0.21{\rm\,GeV}<m_{\mu\mu}^{}<0.224$\,GeV\,  and approximately 30 events were observed.
To obtain a~conservative bound, we assume that all the events in the first bin are statistically
Gaussian and can be attributed to the new particle (either a scalar or a pseudoscalar).
Further assuming uniform acceptance, we obtain at 95\% C.L.
\begin{equation}
{\cal B}(K^+\to\pi^+X) \,\,\lesssim\,\, 8.7 \times 10^{-9} \,\,.
\label{K-bound}
\end{equation}
The NA48 Collaboration collected 6 events for  \,$K_S\to\pi^0\mu^+\mu^-$~\cite{Batley:2004wg},
and none of them have the  214.3-MeV invariant mass required if
they originate from the new particle ${\cal A}$.
Using the $K_S$ flux and the acceptance at low $m_{\mu\mu}^{}$ in Ref.~\cite{Batley:2004wg},
we estimate a single event sensitivity of  \,$\bigl(5.3^{+0.6}_{-0.4}\bigr)\times 10^{-10}$.\,
With no events observed and Poisson statistics, this  translates into the 95\%-C.L. bound
\begin{equation}
{\cal B}(K_S^{}\to\pi^0X) \,\,\lesssim\,\, 1.8\times10^{-9} \,\,. \label{KSbound}
\end{equation}
We employ these bounds when we discuss specific models, but for now we use
the E865 result in Eq.~(\ref{K-bound}), combined with Eq.~(\ref{keq}), to find
\begin{equation}
\left|g_{\cal A}^{}-1.08\, M_{4K}^{}\right| \,\,\lesssim\,\, 4.4\times 10^{-9}  \,\,.
\label{kbound}
\end{equation}

\subsection{$\bm{\Sigma^+\to p{\cal A}}$}

In this case, we need two new dimensionless quantities $A_4$ and  $B_4$ to parameterize
the effect of the four-quark operators, writing the amplitude as
\begin{subequations}  \label{S->pA}
\begin{eqnarray}
{\cal M}(\Sigma^+\to p{\cal A}) &=&
i\bar{p}\left(A_{p\cal A}^{}\,-\,B_{p\cal A}^{}\gamma_5^{}\right)\Sigma^+  \,\,,
\end{eqnarray}
where
\begin{eqnarray}
A_{p\cal A}^{} &=&
g_{\cal A}^{}\,\frac{m_\Sigma^{}-m_N^{}}{v} \,+\, A_4^{}\, \frac{f_\pi^{}}{v} \,\,,
\nonumber \\
B_{p\cal A}^{} &=&
g_{\cal A}^{}\,(D-F)\,\frac{m_\Sigma^{}+m_N^{}}{v}\,\frac{m_K^2}{m_K^2-m_{\cal A}^2}
\,+\,  B_4^{}\, \frac{f_\pi^{}}{v}  \,\,,
\hspace*{1em}
\end{eqnarray}
\end{subequations}
the parameters $D$ and $F$ coming from a chiral Lagrangian to be discussed in a later
section and $\,f_\pi^{}=92.4$\,MeV\, being the pion-decay constant.
The resulting branching ratio is
\begin{eqnarray}
{\cal B}(\Sigma^+ \to p{\cal A}) &=&
1.91\times 10^6 \left|g_{\cal A}^{}+0.36\,A_4^{}\right|^2 \,+\,
4.84\times 10^4 \left|g_{\cal A}^{}+0.14\,B_4^{}\right|^2
\label{gensig}
\end{eqnarray}
with the choice  $\,D-F=0.25$.\,
Combining the statistical and systematic errors of the HyperCP measurement~\cite{Park:2005ek}
in quadrature, we require
\begin{eqnarray}
{\cal B}(\Sigma^+\to p{\cal A}) &=& \bigl(3.1^{+2.8}_{-2.4}\bigr)\times 10^{-8} \,\,,
\label{bound2}
\end{eqnarray}
and therefore
\begin{eqnarray}
\left|g_{\cal A}^{} + 0.36\, A_4^{}\right| &=& (1.3\pm 0.6) \times 10^{-7}  \,\,,
\label{sigbound}
\end{eqnarray}
where we have used the larger of the errors in Eq.~(\ref{bound2}) and  ignored the contribution
from the P-wave term in Eq.~(\ref{gensig}), assuming that  \,$B_4\lesssim A_4$.\,
This assumption is satisfied by all the models we discuss,
but when checking a specific model, we do so without neglecting $B_4$.

A comparison of Eqs.~(\ref{kbound}) and~(\ref{sigbound}) shows why it is not possible to have
a (pseudo)scalar with penguin-like flavor-changing neutral couplings, as in Eq.~(\ref{quark}),
as an explanation for the HyperCP result given the constraints from kaon decay.
It also shows how this is no longer true if there are four-quark contributions to
the amplitudes that are comparable to the penguin amplitudes.
In particular, if we assume that in a given model  $g_{\cal A}^{}$, $M_{4K}$, and $A_4$  have
comparable magnitudes, we see that in order to satisfy both Eqs.~(\ref{kbound})
and~(\ref{sigbound}) we need a cancelation between the two- and four-quark contributions to
the kaon amplitude that reduces them by a factor of about 20.
As we will show in later sections, this is possible in many models.
For this cancelation to work, however, $g_{\cal A}^{}$ and  $M_{4K}$ must also have similar phases.
As we will see, this is a requirement that is much harder to satisfy.
In the simple models we consider in this paper, the phase of $g_{\cal A}^{}$ is much larger
than the phase of $M_{4K}$ so that the cancelation does not happen for the imaginary part.

\subsection{$\bm{b\to s X}$}

Finally we consider the constraints on the new particle from its nonobservation in $B$-meson
decay.
In this case, the four-quark contributions are negligible, and we can neglect $m_s^{}$
compared to $m_b^{}$.
The Lagrangian for  \,$b\to s X$\,  can then be expressed as
\begin{eqnarray}
{\cal L}_{Xbs}^{} &=& \frac{g^\prime\,m_b^{}}{v}\, \bar{s}\bigl(1+\gamma_5^{}\bigr)b\,X
\,\,+\,\, {\rm H.c.}  \,\,,
\label{gbi}
\end{eqnarray}
where  \,$g'=g_{\cal H}^\prime\,(ig_{\cal A}^\prime)$\,  for  \,$X=\cal H\,(\cal A)$.\,
This leads to the partial decay rate
\begin{eqnarray}
\Gamma(b\to s X) &\simeq& |g^\prime|^2\,\frac{m_b^3}{8\pi v^2}  \,\,.
\label{gambx}
\end{eqnarray}
Using for illustration  \,$m_b^{}=4.3$\,GeV\,  and the  $B^+$ lifetime~\cite{hfag} results in
\begin{eqnarray}
{\cal B}(b\to s X) &=& 1.3\times 10^8\, |g^\prime|^2  \,\,.
\label{bxrate}
\end{eqnarray}
One could obtain a similar number for  \,$b\to d X$.\,

The latest experimental average
\,${\cal B}(b\to s\mu^+\mu^-)=\bigl(4.27^{+1.23}_{-1.22}\bigr)\times 10^{-6}$~\cite{hfag}
covers the full kinematic range for  $m_{\mu\mu}^{}$.
To constrain  $g^\prime$, it is better to limit the comparison to the
measured rate at the lowest measured $m_{\mu\mu}^{}$ invariant-mass bin.
BABAR quotes in Table II of Ref.~\cite{Aubert:2004it}
\begin{eqnarray}
{\cal B}(b\to s\ell^+\ell^-)_{m_{\ell^+\ell^-}\in[0.2{\rm\,GeV},1.0{\rm\,GeV}]}^{}  &=&
\bigl(0.08\pm 0.36^{+0.07}_{-0.04}\bigr)\times 10^{-6}  \,\,.
\end{eqnarray}
This is an average for electrons and muons, but no noticeable difference between them was found.
Belle quotes on Table IV of Ref.~\cite{Iwasaki:2005sy} the corresponding number
\begin{eqnarray}
{\cal B}(b\to s\ell^+\ell^-)_{m_{\ell^+\ell^-}\in[0.2{\rm\,GeV},1.0{\rm\,GeV}]}^{} &=&
\bigl(11.3\pm 4.8^{+4.6}_{-2.7}\bigr)\times 10^{-7}  \,\,.
\end{eqnarray}
To be conservative, we constrain the Higgs coupling by requiring that the induced rate be below
the 95\%-C.L. upper-range of the measured  \,$b\to s\ell^+\ell^-$\,  rate in the lowest
measured $m_{\mu\mu}^{}$ bin.
Thus, combining errors in quadrature for the more restrictive BABAR result gives
\begin{eqnarray}
{\cal B}(b\to s\ell^+\ell^-)_{m_{\ell^+\ell^-}<1\rm\,GeV}^{} &\lesssim& 8.0\times 10^{-7}
\hspace{2em} \rm (BABAR)
\end{eqnarray}
and correspondingly
\begin{eqnarray}
|g^\prime| &\lesssim& 7.8\times 10^{-8}  \,\,.\label{ggbs}
\end{eqnarray}
The exclusive  \,$B\to(K,K^\star)\mu^+\mu^-$\,  modes have been measured, but the resulting
constraints are not better than Eq.~(\ref{ggbs}). This constraint, Eq.~(\ref{ggbs}), is
difficult to satisfy in models where $g_{\cal A}^{}$ and $g^\prime$ are related by top-quark
CKM angles, as happens in the simple models we consider here.

\section{Scalar Higgs boson\label{scalarH}}

In this section we discuss in detail the case of a light Higgs boson in the standard model
and in the two-Higgs-doublet model.
We will use known low-energy theorems to implement the four-quark contributions to kaon and
hyperon amplitudes.

\subsection{Two-quark $\bm{|\Delta S|=1}$  interactions}

The effective Lagrangian for the  $sd\cal H$  coupling, where $\cal H$ is either
the standard-model Higgs boson $H^0$ or the lightest scalar Higgs-boson $h^0$ in the 2HDM,
has been much discussed in the literature~\cite{sdh_sm,sdh_2hdm,sdH,Leutwyler:1989xj,Dawson:1989bm} and
can be written as  ${\cal L}_{{\cal H}sd}$  in Eq.~(\ref{Hsd}), where
\begin{eqnarray}   \label{gH}
g_{\cal H}^{}  \,\,=\,\,
\frac{G_F^{}}{4\sqrt2\,\pi^2}\sum_{q=u,c,t}m_q^2 V^*_{qd}V_{qs}^{}\,F(q)  \,\,,
\end{eqnarray}
with  $V_{kl}^{}$ being the elements of the Cabibbo-Kobayashi-Maskawa (CKM) matrix and
$F(q)$ depending on the model.
In the SM, for a Higgs mass much smaller than the $W$ mass,
\begin{eqnarray}
F(q)  \,\,=\,\,  3/4   \,\,,
\end{eqnarray}
whereas in the 2HDM the expression for $F(q)$ is much lengthier~\cite{sdh_2hdm,Dawson:1989bm}.

Using CKM and mass parameters from Ref.~\cite{Charles:2004jd}, we find in the SM
\begin{eqnarray}
g_{\cal H}^{}  &=&  (-1.3-0.6i)\times10^{-6}  \,\,,
\end{eqnarray}
to be compared with  Eqs.~(\ref{kbound}) and~(\ref{sigbound})  above.
Employing the expression for  $F(q)$ derived in Ref.~\cite{Dawson:1989bm},
we obtain a similar number in the 2HDM type II, for instance,
\begin{eqnarray}  \label{gH2hdm}
g_{\cal H}^{}  &=&  (5.0+1.9i)\times10^{-7}
\end{eqnarray}
for the parameters
\begin{eqnarray}  \label{2hdmpar}
\tan\beta \,\,\simeq\,\, 2.57 \,\,, \hspace{2em}  \sin(\beta-\alpha) \,\,\simeq\,\, 0.149  \,\,,
\hspace{2em}   m_{H^+}^{} \,\,=\,\, 250{\rm~GeV}  \,\,,
\end{eqnarray}
where  \,$\tan\beta$\,  is the ratio of vacuum expectation values of the two Higgs doublets,
$\alpha$ the mixing angle in the neutral-Higgs-boson mass matrix, and  $m_{H^+}$
the mass of the charged Higgs bosons.\footnote{We have also set \,$\kappa=m_{H^+}^2/m_W^2$\,
in  $F(q)$,  where  $\kappa$ is defined in Ref.~\cite{Dawson:1989bm}.}
We note that the $\alpha$ and $\beta$ values above satisfy the constraint
\,$\sin^2(\beta-\alpha)<0.06$\,  from LEP~\cite{Abbiendi:2004gn}.
We see right away that $g_{\cal H}^{}$ can be in the right ball park to explain the HyperCP
observation, Eq.~(\ref{sigbound}), but conflicts with the kaon bound, Eq.~(\ref{kbound}).

To evaluate the hadronic amplitudes from this 2-quark contribution, we employ chiral
perturbation theory.
Using the operator matching of Ref.~\cite{He:2005we}, we write the lowest-order chiral
realization of  ${\cal L}_{{\cal H}sd}$  as
\begin{eqnarray}   \label{LsdH}
{\cal L}_{\cal H}^{}  &=&
b_D^{} \left\langle \bar{B}{}^{} \left\{ h_{\cal H}^{}, B^{} \right\} \right\rangle
+ b_F^{} \left\langle \bar{B}{}^{} \left[ h_{\cal H}^{}, B^{} \right] \right\rangle
+ b_0^{} \left\langle h_{\cal H}^{} \right\rangle \left\langle\bar{B}{}^{}B^{}\right\rangle
\,+\,  \mbox{$\frac{1}{2}$} f^2 B_0^{} \left\langle h_{\cal H}^{} \right\rangle
\,\,+\,\,  {\rm H.c.}    \,\,,
\end{eqnarray}
where  $\,\langle\cdots\rangle\equiv{\rm Tr}(\cdots)\,$  in flavor-SU(3) space,
$\,f=f_\pi^{}=92.4\rm\,MeV$,\,  and
\begin{eqnarray}
h_{\cal H}^{}  \,\,=\,\,
-2g_{\cal H}^{}\,\bigl(\xi^\dagger h M\xi^\dagger+\xi M h\xi\bigr)\frac{\cal H}{v} \;,
\end{eqnarray}
with  $h$ being a 3$\times$3-matrix  having elements
$\,h_{kl}^{}=\delta_{k2}^{}\delta_{3l}^{}\,$  which selects out  $\,s\to d\,$  transitions,
$\,M={\rm diag}(\hat{m},\hat{m},m_s^{})=
{\rm diag}\bigl(m_\pi^2,m_\pi^2,2m_K^2-m_\pi^2\bigr)/(2B_0^{})\,$  the quark-mass matrix
in the isospin-symmetric limit  $\,m_u^{}=m_d^{}=\hat{m}$,\,
and the baryon and meson fields represented by the usual 3$\times$3-matrices
$B$ and  \,$\Sigma=\xi\xi=e^{i\varphi/f}$,\, respectively.

To derive amplitudes, we also need the chiral Lagrangian for the strong interactions of
the hadrons~\cite{Gasser:1983yg,bcpt}.
At leading order in the derivative and $m_s^{}$ expansions, it can be written as
\begin{eqnarray}   \label{Lstrong}
{\cal L}_{\rm s}^{}  &=&
\left\langle \bar{B}^{}\, {i}\gamma^\mu \bigl(
\partial_\mu^{}B+\bigl[{\cal V}_\mu^{},B \bigr] \bigr) \right\rangle
- m_0^{} \left\langle \bar{B}^{} B^{} \right\rangle
+ D \left\langle \bar{B}^{} \gamma^\mu\gamma_5^{}
 \left\{ {\cal A}_\mu^{}, B^{} \right\} \right\rangle
+ F \left\langle \bar{B}^{} \gamma^\mu\gamma_5^{}
 \left[ {\cal A}_\mu^{}, B^{} \right] \right\rangle
\nonumber \\ && +\,\,
b_D^{} \left\langle\bar{B}^{}\left\{M_+,B^{}\right\}\right\rangle
+ b_F^{} \left\langle\bar{B}^{}\left[M_+,B^{}\right]\right\rangle
+ b_0^{} \left\langle M_+ \right\rangle \left\langle \bar{B}^{} B^{} \right\rangle
\nonumber \\ &&
+\,\,
\mbox{$\frac{1}{4}$} f^2 \left\langle
\partial^\mu\Sigma^\dagger\, \partial_\mu^{}\Sigma \right\rangle +
\mbox{$\frac{1}{2}$} f^2 B_0^{} \left\langle M_+ \right\rangle  \,\,,
\end{eqnarray}
where
$\,{\cal V}^\mu =
\frac{1}{2}\bigl(\xi\,\partial^\mu\xi^\dagger+\xi^\dagger\,\partial^\mu\xi\bigr),\,$
$m_0^{}$ is the baryon mass in the chiral limit,
$\,{\cal A}^\mu  = \frac{i}{2}
\bigl(\xi\,\partial^\mu\xi^\dagger-\xi^\dagger\,\partial^\mu\xi\bigr),\,$
and  \,$M_+^{}=\xi^\dagger M\xi^\dagger+\xi M^\dagger\xi$,\,
with  further details being given in Ref.~\cite{He:2005we}.

\begin{figure}[t]
\begin{picture}(160,100)(-80,-20)
\Text(-32,0)[r]{\footnotesize$\Sigma^+$} \Line(-30,0)(0,0)
\Line(0,0)(0,30) \Text(0,35)[]{\footnotesize$\cal H$}
\Line(0,0)(30,0) \Text(32,0)[l]{\footnotesize$p$}
\SetWidth{1} \BBoxc(0,0)(4,4) \Text(0,-20)[]{\footnotesize(a)}
\end{picture}
\begin{picture}(160,100)(-80,-20)
\Text(-32,0)[r]{\footnotesize$\Sigma^+$} \Line(-30,0)(0,0)
\Line(0,0)(0,50) \Text(2,15)[l]{\footnotesize$\bar{K}^0$}
\Text(0,55)[]{\footnotesize$\cal H$} \Line(0,0)(30,0) \Text(32,0)[l]{\footnotesize$p$}
\Vertex(0,0){2} \SetWidth{1} \BBoxc(0,30)(4,4) \Text(0,-20)[]{\footnotesize(b)}
\end{picture}
\caption{\label{sdHdiagrams}%
Diagrams contributing to  $\,\Sigma^+\to p\cal H$\,  arising from  ${\cal L}_{{\cal H}sd}$
at leading order in  $\chi$PT.
The square vertices come from  ${\cal L}_{\cal H}$  in  Eq.~(\ref{LsdH}),
and the solid vertex from  ${\cal L}_{\rm s}$  in  Eq.~(\ref{Lstrong}).
}
\end{figure}
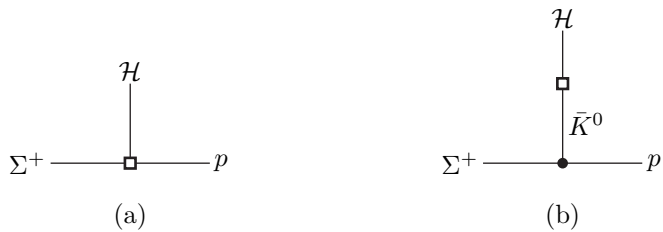

 {From}  ${\cal L}_{\cal H}$ and  ${\cal L}_{\rm s}$, we derive the leading-order diagrams
shown in Fig.~\ref{sdHdiagrams} for  $\,\Sigma^+\to p\cal H$,\, yielding the amplitude
\begin{eqnarray}   \label{MSpH}
{\cal M}_{2q}^{}(\Sigma^+\to p{\cal H})  &=&
g_{\cal H}^{}\,\frac{m_\Sigma^{}-m_N^{}}{v}\,\frac{m_K^2}{m_K^2-m_\pi^2}\,\bar{p}\Sigma^+
\nonumber \\ && -\,\,
g_{\cal H}^{}\,(D-F)\,\frac{m_\Sigma^{}+m_N^{}}{v}\,\frac{m_K^2-m_\pi^2}{m_K^2-m_{\cal H}^2}\,
\bar{p}\gamma_5^{}\Sigma^+     \,\,,
\end{eqnarray}
where the two terms correspond to the two diagrams, respectively,  $m_{\Sigma,N}^{}$  are
isospin-symmetric masses, and we have used the relations
\,$m_\Sigma^{}-m_N^{}=2\bigl(b_D^{}-b_F^{}\bigr)\bigl(m_s^{}-\hat{m}\bigr),\,$
\,$m_K^2=B_0^{}\bigl(\hat{m}+m_s^{}\bigr),\,$  and  \,$m_\pi^2=2B_0^{}\hat{m}$\,
derived from  ${\cal L}_{\rm s}$.
Numerically, we will allow $D$ and $F$ to have the ranges
\,$0.6\le D\le 0.8$\,  and  \,$0.4\le F\le0.5$\,~\cite{bcpt},  leading to
\begin{eqnarray}  \label{d-f}
0.1 \,\le\, D-F \,\le\, 0.4   \;,
\end{eqnarray}
which is their combination occurring in our amplitudes.

It follows that the contribution of  ${\cal L}_{{\cal H}sd}$  to the branching ratio of
$\,\Sigma^+\to p\cal H$\, for \,$m_{\cal H}^{}=214.3$\,MeV\, and the middle value
$\,D-F=0.25$\, is in the SM
\begin{equation}
{\cal B}_{2q}^{}(\Sigma^+\to p{\cal H}) \,\,=\,\, (40 + 1) \times 10^{-7}  \,\,,
\end{equation}
where we have ignored the imaginary ($CP$ violating) part of the amplitude, and
the two numbers correspond to the contributions from the scalar and pseudoscalar
flavor-changing couplings, respectively.
Evidently, the scalar contribution is much larger than what HyperCP saw, but the pseudoscalar
contribution is within the range.
This, however, is only part of the story, as there are in addition 4-quark contributions
to be discussed in the next subsection.

Also from  ${\cal L}_{\cal H}$,  we derive the leading-order diagram for
\,$K\to\pi\cal H$,\, which is that in Fig.~\ref{sdHdiagrams}(a) with  $\Sigma^+$ ($p$)
replaced by  $K$ ($\pi$) and arises from the scalar coupling in  ${\cal L}_{{\cal H}sd}$.
The resulting amplitude is
\begin{eqnarray}   \label{M2q[K->pi-H]}
{\cal M}_{2q}^{}(K^+\to\pi^+{\cal H}) \,\,=\,\,
-\sqrt2\,{\cal M}_{2q}^{}(K^0\to\pi^0{\cal H})
\,\,=\,\,  \frac{-g_{\cal H}^{}\, m_K^2}{v}  \,\,,
\end{eqnarray}
and so  \,${\cal M}_{2q}(K_L^{}\to\pi^0{\cal H})=
-{\rm Re}\,{\cal M}_{2q}(K^+\to\pi^+{\cal H})$.\,
Dropping again the imaginary parts of the amplitudes, we obtain in the SM the branching ratios
\begin{eqnarray}
{\cal B}_{2q}^{}(K^+\to\pi^+{\cal H})  \,\,=\,\, 9.3\times10^{-4}   \,\,, \hspace{2em}
{\cal B}_{2q}^{}(K_L^{}\to\pi^0{\cal H})  \,\,=\,\, 3.9\times10^{-3}   \,\,.
\end{eqnarray}
These numbers would easily be incompatible with that in Eq.~(\ref{K-bound})
and the 95\%-C.L. bound\footnote{We have inferred this number from Ref.~\cite{Alavi-Harati:2000hs}
which reported  \,${\cal B}(K_L\to\pi^0\mu^+\mu^-)<3.8\times10^{-10}$\,  at 90\% C.L.}
\begin{eqnarray}  \label{KLbound}
{\cal B}(K_L^{}\to\pi^0\mu^+\mu^-)  &<& 4.9\times10^{-10}   \,\,,
\end{eqnarray}
but, as in the $\Sigma^+$ case, there are 4-quark contributions that
have to be considered as well.

The situation is similar in the 2HDM.
Adopting the real part of the coupling in Eq.~(\ref{gH2hdm}), for example, we find
\begin{eqnarray}
{\cal B}_{2q}^{}(\Sigma^+\to p{\cal H}) &=& (56 + 1) \times 10^{-8}  \,\,,
\nonumber \\ \vphantom{\sum}
{\cal B}_{2q}^{}(K^+\to\pi^+{\cal H})  \,\,=\,\, 1.3\times10^{-4}   \,\,, &&
{\cal B}_{2q}^{}(K_L^{}\to\pi^0{\cal H})  \,\,=\,\, 5.4\times10^{-4}   \,\,.
\end{eqnarray}

\subsection{Four-quark $\bm{|\Delta S|=1}$ interactions}

The hadronic interactions of a light Higgs-boson due to 4-quark  \,$|\Delta S|=1$\, operators
are best accounted for in the chiral Lagrangian approach.
The dominant contribution is generated by the  $\,|\Delta I|=\frac{1}{2}\,$ component of
the effective Hamiltonian transforming as $(8_{\rm L}^{},1_{\rm R}^{})$.
The corresponding Lagrangian at leading order is given by~\cite{bcpt,dgh}.
\begin{eqnarray} \label{Lweak}
{\cal L}_{\rm w}^{}  &=&
h_D^{} \left\langle \bar B \left\{ \xi^\dagger h \xi\,,\,B \right\} \right\rangle
+ h_F^{} \left\langle \bar B \left[ \xi^\dagger h \xi\,,\,B \right] \right\rangle
\nonumber \\ && +\,\,
\gamma_8^{}f^2 \left\langle h\,\partial_\mu^{}\Sigma\,
\partial^\mu \Sigma^\dagger \right\rangle
+ 2\tilde{\gamma}_8^{} f^2 B_0^{} \left\langle h \xi M_+ \xi^\dagger \right\rangle
\,\,+\,\,  {\rm H.c.}   \,\,,
\end{eqnarray}
where  $h_{D,F}^{}$  can be extracted from hyperon nonleptonic decays,
$\,\gamma_8^{}=-7.8\times10^{-8}$\,  from  $\,K\to\pi\pi$,\,  the sign following from various
predictions~\cite{Gunion:1989we,sdH,Leutwyler:1989xj,Bijnens:1998ee},  and  $\tilde{\gamma}_8^{}$
is unknown as it does not contribute to any process with only kaons and pions.

The 4-quark \,$|\Delta S|=1$\, interactions of a light Higgs-boson $\cal H$ arises from
its tree-level couplings to quarks and $W^\pm$ bosons, as well as from its coupling to
gluons induced by a triangle diagram with heavy quarks in the loop.
To obtain the relevant chiral Lagrangians, one starts with  ${\cal L}_{\rm s,w}$  above and
follows the prescription given in Refs.~\cite{sdH,Leutwyler:1989xj,Dawson:1989bm,Gunion:1989we}.
The results are
\begin{eqnarray}   \label{LsH}
{\cal L}_{\rm s}^{\cal H}  &=&
\left( \mbox{$\frac{1}{4}$}\, c_1^{}\,  f^2 \left\langle
\partial^\mu\Sigma^\dagger\, \partial_\mu^{}\Sigma \right\rangle \vphantom{|_|^|}
+ \mbox{$\frac{1}{2}$}\, c_2^{}\, f^2 B_0^{} \left\langle M_+ \right\rangle
+ \mbox{$\frac{1}{2}$}\,f^2 B_0^{}\,\bigl\langle\hat{M}_+-M_+\bigr\rangle\right)\frac{\cal H}{v}
\,-\,  k_1^{}\, m_0^{} \left\langle \bar{B}^{} B^{} \right\rangle \frac{\cal H}{v}
\nonumber \\ && +\,\,
k_2^{} \left( b_D^{}\, \bigl\langle \bar{B}^{}\, \bigl\{ \hat{M}_+, B^{} \bigr\} \bigr\rangle
+ b_F^{}\, \bigl\langle \bar{B}^{}\, \bigl[ \hat{M}_+, B^{} \bigr] \bigr\rangle
+ b_0^{}\, \bigl\langle \hat{M}_+ \bigr\rangle\, \bigl\langle\bar{B}B\bigr\rangle \right)
\frac{\cal H}{v} \,\,,
\end{eqnarray}
\begin{eqnarray} \label{LwH}
{\cal L}_{\rm w}^{\cal H}  &=&
\left[ \gamma_8^{}\, c_3^{}\, f^2 \left\langle h\,\partial_\mu^{}\Sigma\,
\partial^\mu \Sigma^\dagger \right\rangle
+ 2\tilde{\gamma}_8^{}\, c_4^{}\, f^2 B_0^{}
\left\langle h \xi M_+ \xi^\dagger \right\rangle
+ 2\tilde{\gamma}_8^{}\, f^2 B_0^{}\,
\bigl\langle h \xi \bigl(\hat{M}_+-M_+\bigr) \xi^\dagger \bigr\rangle \right] \frac{\cal H}{v}
\nonumber \\ && +\,\,
k_3^{} \left( h_D^{}\left\langle \bar B\left\{\xi^\dagger h \xi\,,\,B \right\} \right\rangle
+ h_F^{}\left\langle\bar B\left[\xi^\dagger h\xi\,,\,B\right]\right\rangle\right)\frac{\cal H}{v}
\,\,+\,\,  {\rm H.c.}   \,\,,
\end{eqnarray}
where
\begin{eqnarray}
\begin{array}{c}   \displaystyle
c_1^{}  \,\,=\,\, 2k_G^{}  \,\,,  \hspace{2em}
c_2^{}  \,\,=\,\, 3k_G^{}+1  \,\,,  \hspace{2em}
c_3^{}  \,\,=\,\, 4k_G^{}-2k_W^{}  \,\,,  \hspace{2em}
c_4^{}  \,\,=\,\, 5k_G^{}-2k_W^{}+1  \,\,,
\vspace{1ex} \\   \displaystyle
k_1^{}  \,\,=\,\, k_G^{}  \,\,,  \hspace{2em}
k_2^{}  \,\,=\,\, 1  \,\,,  \hspace{2em}
k_3^{}  \,\,=\,\, 3k_G^{}-2k_W^{}  \,\,,
\vspace{1ex} \\   \displaystyle
\hat{M}_+^{} \,\,=\,\, \xi^\dagger\hat{M}\xi^\dagger+\xi\hat{M}^\dagger\xi  \,\,,
\end{array}
\end{eqnarray}
with
\begin{eqnarray}
k_G^{}  \,\,=\,\,  \frac{2(2k_u^{}+k_d^{})}{27}    \,\,,  \hspace{2em}
\hat{M}  \,\,=\,\,  {\rm diag}\bigl(k_u^{}\hat{m},\,k_d^{}\hat{m},\,k_d^{}m_s^{}\bigr)  \,\,,
\end{eqnarray}
the expression for $k_G^{}$ corresponding to 3 heavy and 3 light quarks.
The parameters  $k_{u,d}^{}$, $k_W^{}$, and $k_G^{}$  come from the couplings of
$\cal H$ to light quarks, $W^\pm$, and the gluons, respectively,
and depend on the model of the Higgs sector.
Thus
\begin{eqnarray}  \label{ksm}
k_u^{}  \,\,=\,\, k_d^{}  \,\,=\,\, k_W^{} &=& 1  \hspace{3em} \mbox{in the SM}   \,\,,
\\ \vphantom{\int}
\label{kI}
k_u^{}  \,\,=\,\, k_d^{}  \,\,=\,\,  \frac{\cos\alpha}{\sin\beta}  \,\,,  \hspace{2em}
k_W^{}  &=&  \sin(\beta-\alpha)  \hspace{3em}  \mbox{in the 2HDM~I} \,\,,
\\
\label{kII}
k_u^{}  \,\,=\,\, \frac{\cos\alpha}{\sin\beta}  \,\,,  \hspace{2em}
k_d^{}  \,\,=\,\,  -\frac{\sin\alpha}{\cos\beta}  \,\,,  &&
k_W^{}  \,\,=\,\,  \sin(\beta-\alpha)  \hspace{3em}  \mbox{in the 2HDM~II} \,\,.
\end{eqnarray}
The parameters  $c_{1,2,3,4}^{}$  for the meson terms have already been obtained
in the literature~\cite{Gunion:1989we,sdH,Leutwyler:1989xj,Dawson:1989bm,Prades:1990vn},
whereas the new ones  $k_{1,2,3}^{}$  follow from how the baryon parameters
depend on masses:
\,$m_0^{}\sim\Lambda$,\,  $\,b_{D,F,0}^{}\sim 1$,\,  \,$\chi_+^{}\sim\Lambda m_q^{}$,\,
and  \,$h_{D,F}^{}\sim\Lambda^3/m_W^2$,\,  where $\Lambda$ is a QCD mass scale.
Note that we work in that basis in which the mass terms in the Lagrangians are
not diagonal and must therefore include the corresponding tadpole diagrams in our
calculation.

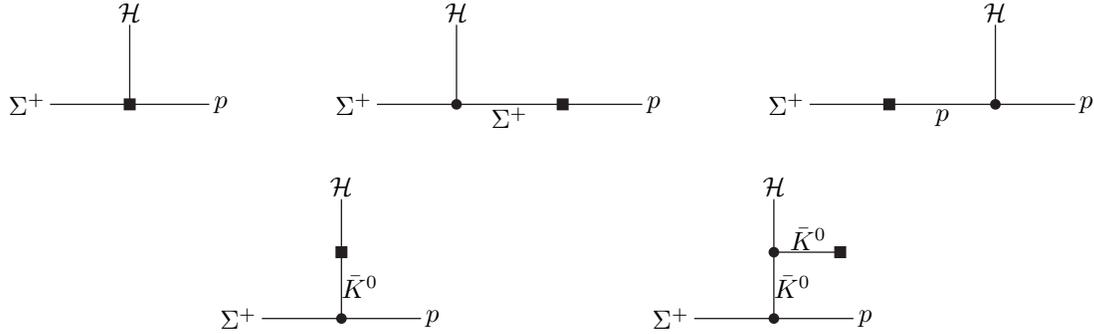
\begin{figure}[t]
\begin{picture}(120,80)(-60,-20)
\Text(-32,0)[r]{\footnotesize$\Sigma^+$} \Line(-30,0)(30,0)
\Line(0,0)(0,30) \Text(0,35)[]{\footnotesize$\cal H$}
\Text(32,0)[l]{\footnotesize$p$} \CBoxc(0,0)(4,4){Black}{Black}
\end{picture}
\begin{picture}(160,80)(-80,-20)
\Text(-52,0)[r]{\footnotesize$\Sigma^+$} \Line(-50,0)(50,0)
\Line(-20,0)(-20,30) \Text(-20,35)[]{\footnotesize$\cal H$}
\Text(0,-5)[]{\footnotesize$\Sigma^+$} \Vertex(-20,0){2}
\Text(52,0)[l]{\footnotesize$p$} \CBoxc(20,0)(4,4){Black}{Black}
\end{picture}
\begin{picture}(160,80)(-80,-20)
\Text(-52,0)[r]{\footnotesize$\Sigma^+$} \Line(-50,0)(50,0)
\Line(20,0)(20,30) \Text(20,35)[]{\footnotesize$\cal H$}
\Text(0,-5)[]{\footnotesize$p$} \Vertex(20,0){2}
\Text(52,0)[l]{\footnotesize$p$} \CBoxc(-20,0)(4,4){Black}{Black}
\end{picture}
\\
\begin{picture}(160,70)(-80,-10)
\Text(-32,0)[r]{\footnotesize$\Sigma^+$} \Line(-30,0)(30,0) \Line(0,0)(0,45)
\Text(0,50)[]{\footnotesize$\cal H$} \Vertex(0,0){2} \CBoxc(0,25)(4,4){Black}{Black}
\Text(32,0)[l]{\footnotesize$p$} \Text(7,12)[]{\footnotesize$\bar{K}^0$}
\end{picture}
\begin{picture}(160,70)(-80,-10)
\Text(-32,0)[r]{\footnotesize$\Sigma^+$} \Line(-30,0)(30,0) \Line(0,0)(0,45)
\Line(0,25)(25,25) \Text(0,50)[]{\footnotesize$\cal H$} \Vertex(0,0){2} \Vertex(0,25){2}
\CBoxc(25,25)(4,4){Black}{Black} \Text(32,0)[l]{\footnotesize$p$}
\Text(7,12)[]{\footnotesize$\bar{K}^0$} \Text(13,30)[]{\footnotesize$\bar{K}^0$}
\end{picture}  \vspace*{-1ex}
\caption{\label{4qHdiagrams}%
Diagrams contributing to  $\,\Sigma^+\to p\cal H$\,  arising from  the 4-quark operators.
The square vertices come from  ${\cal L}_{\rm w}^{({\cal H})}$  in  Eqs.~(\ref{Lweak})
and~(\ref{LwH}), whereas the dots are from  ${\cal L}_{\rm s}^{({\cal H})}$  in
Eqs.~(\ref{Lstrong}) and~(\ref{LsH}).
}
\end{figure}

For \,$\Sigma^+\to p\cal H$,\, we derive from  ${\cal L}_{\rm s,w}^{({\cal H})}$
the diagrams shown in Fig.~\ref{4qHdiagrams}, finding
\begin{eqnarray}   \label{MSpH'}
{\cal M}_{4q}^{}(\Sigma^+\to p{\cal H})  &=&
(k_d^{}-3k_G^{}+2k_W^{})\frac{h_D^{}-h_F^{}}{v}\, \bar{p}\Sigma^+
\nonumber \\ && +\,\,
4(k_G^{}-k_W^{})\,(D-F)\,\tilde{\gamma}_8^{}\,\frac{m_\Sigma^{}+m_N^{}}{v}\,
\frac{m_K^2-m_\pi^2}{m_K^2-m_{\cal H}^2}\, \bar{p}\gamma_5^{}\Sigma^+  \,\,,
\end{eqnarray}
where the first term comes from the upper three diagrams, which are at leading order, and
the second term results from the lower two diagrams, which are at next-to-leading order.
Now, the combination  $\,h_D$$-$$h_F$\,  also occurs in the amplitude for
\,$\Sigma^+\to p\pi^0$,\,  which we write as
\begin{eqnarray}
{\cal M}(\Sigma^+\to p\pi^0)  \,\,=\,\,
{i}\bar{p}\, \bigl(A_{p\pi^0}^{}-B_{p\pi^0}^{}\gamma_5^{}\bigr)\,\Sigma^+   \;,
\end{eqnarray}
where from  ${\cal L}_{\rm s,w}$
\begin{eqnarray}  \label{ABt}
A_{p\pi^0}^{} \,\,=\,\, \frac{-h_D^{}+h_F^{}}{2\,f}  \,\,, \hspace{2em}
B_{p\pi^0}^{} \,\,=\,\,
(D-F)\frac{h_D^{}-h_F^{}}{2\,f}\,\,\frac{m_\Sigma^{}+m_N^{}}{m_\Sigma^{}-m_N^{}}  \,\,.
\end{eqnarray}
Since from experiment~\cite{pdg}
\begin{eqnarray}  \label{ABx}
A_{p\pi^0}^{}  \,\,=\,\,  -3.25\times10^{-7}  \;,  \hspace*{3em}
B_{p\pi^0}^{}  \,\,=\,\,  26.67\times10^{-7}   \;,
\end{eqnarray}
up to an overall sign, in our numerical evaluation of the 4-quark contributions to
\,$\Sigma^+\to p\cal H$\,  we will explore different  \,$h_D$$-$$h_F$\,  values accordingly.

We can also derive from  ${\cal L}_{\rm s,w}^{({\cal H})}$  the corresponding leading-order
diagrams for  $\,K\to\pi\cal H$,\, which are the upper three in Fig.~\ref{4qHdiagrams}
with $\Sigma^+$ ($p$) replaced by  $K$ ($\pi$) and yield
\begin{eqnarray}   \label{M4q[K->pi-H]}
{\cal M}_{4q}^{}\bigl(K^+\to\pi^+{\cal H}\bigr)  &=&
\frac{\gamma_8^{}}{v}\bigl[ 2(k_W^{}-k_G^{})\bigl(m_K^2+m_\pi^2-m_{\cal H}^2\bigr)
+ (k_d^{}-k_u^{})\,m_\pi^2 \bigr]
\nonumber \\ && +\,\,
\frac{\tilde{\gamma}_8^{}}{v}\,4(k_G^{}-k_W^{})\,m_K^2  \,\,,
\end{eqnarray}
\begin{eqnarray}   \label{M4q[K->pi0H]}
{\cal M}_{4q}^{}\bigl(K^0\to\pi^0{\cal H}\bigr)  &=&
\frac{\gamma_8^{}}{\sqrt2\,v} \Biggl[
2(k_G^{}-k_W^{})\bigl(m_K^2+m_\pi^2-m_{\cal H}^2\bigr)
+ (k_u^{}-k_d^{})\frac{m_\pi^2\,m_K^2}{m_K^2-m_\pi^2} \Biggr]
\nonumber \\ && +\,\,
\frac{\tilde{\gamma}_8^{}}{\sqrt2\,v} \Biggl[ 4(k_W^{}-k_G^{})\,m_K^2
+ (k_d^{}-k_u^{})\frac{m_\pi^2\,m_K^2}{m_K^2-m_\pi^2} \Biggr] \,\,.
\end{eqnarray}
Since  $\tilde{\gamma}_8^{}$  is unknown, in evaluating its effect on
$\,K^+\to\pi^+\cal H$\,  we will allow it to vary from  $-10$ to 10 times $\gamma_8^{}$.
Naively we would expect $\gamma_8^{}$ and $\tilde{\gamma}_8^{}$ to be of the same order.

\subsection{Total contributions}

\begin{figure}[t] \vspace{4ex}
\includegraphics[width=4in]{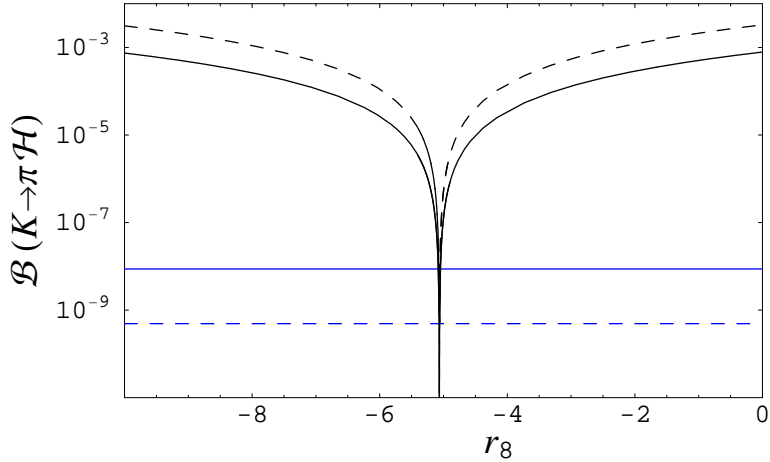}   \vspace{-1ex}
\caption{\label{br(K->piH)}%
Contributions of real parts of total amplitudes for  $\,K^+\to\pi^+\cal H$\, (solid curve)
and $\,K_L\to\pi^0\cal H$\, (dashed curve) in the SM to their branching ratios as functions
of  \,$r_8^{}=\tilde{\gamma}_8^{}/\gamma_8^{}$\,  for  $\,m_{\cal H}^{}=214.3$\,MeV.\,
The horizontal lines are the corresponding upper bounds in Eqs.~(\ref{K-bound}) and~(\ref{KLbound}).
}\end{figure}

The total amplitude for  \,$K^+\to\pi^+\cal H$\,  comes from the sum of the contributions
in  Eqs.~(\ref{M2q[K->pi-H]}) and~(\ref{M4q[K->pi-H]}).
If the $CP$-violating terms in the amplitudes are ignored, it is possible for the
2-quark and 4-quark contributions to cancel.
We show this possibility in Fig.~\ref{br(K->piH)}, where we plot the resulting
branching ratio in the SM as a function of the ratio
$\,r_8^{}\equiv\tilde{\gamma}_8^{}/\gamma_8^{}$\,  for  $\,m_{\cal H}^{}=214.3$\,MeV.\,
We find that  \,${\cal B}(K^+\to\pi^+{\cal H})=0$\,  when  $\,r_8^{}\simeq-5.1$\,  and that,
as the figure indicates, for only a very narrow range of $r_8^{}$ around this value does
the branching ratio ever fall below the upper limit in Eq.~(\ref{K-bound}).
In Fig.~\ref{br(K->piH)}, we also plot the corresponding branching ratio of
the isospin-related mode  \,$K_L\to\pi^0\cal H$.\,

\begin{figure}[ht]
\vspace{4ex}
\includegraphics[width=4in]{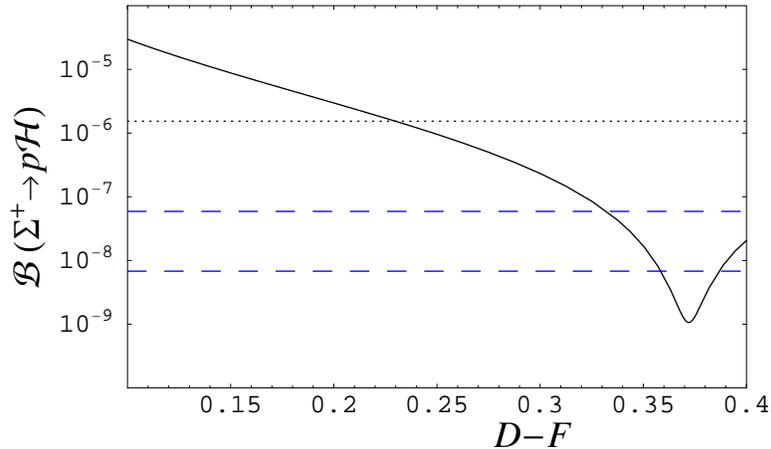}   \vspace{-1ex}
\caption{\label{br(S->pH)}%
Contribution of real part of total amplitude for  \,$\Sigma^+\to p\cal H$\, to its branching
ratio in the SM as function of  \,$D$$-$$F$\,  for  $\,m_{\cal H}^{}=214.3$\,MeV\,  and
$\,r_8^{}\simeq-5.1$.\,
The solid (dotted) curve corresponds to  \,$h_D$$-$$h_F$\, extracted from the
P-wave (S-wave) fit to the  $\,\Sigma^+\to p\pi^0$\,  data using Eqs.~(\ref{ABt})
and~(\ref{ABx}).
The dashed lines correspond to the upper and lower bounds in the HyperCP result.
}
\end{figure}

For  \,$\Sigma^+\to p\cal H$,\,  the total amplitude results from adding the contributions
in Eqs.~(\ref{MSpH}) and~(\ref{MSpH'}).
Including only the real part of amplitudes again, and using  $\,r_8^{}\simeq-5.1$\,
determined above,  we plot in Fig.~\ref{br(S->pH)} the branching ratio in the SM as a function
of  \,$D$$-$$F$\,  for the range in Eq.~(\ref{d-f}).
This figure shows that the curve resulting from the P-wave fit using Eqs.~(\ref{ABt})
and~(\ref{ABx}) satisfies the HyperCP constraints for certain $D$$-$$F$ values.


\begin{figure}[t]
\vspace{4ex}
\includegraphics[width=4in]{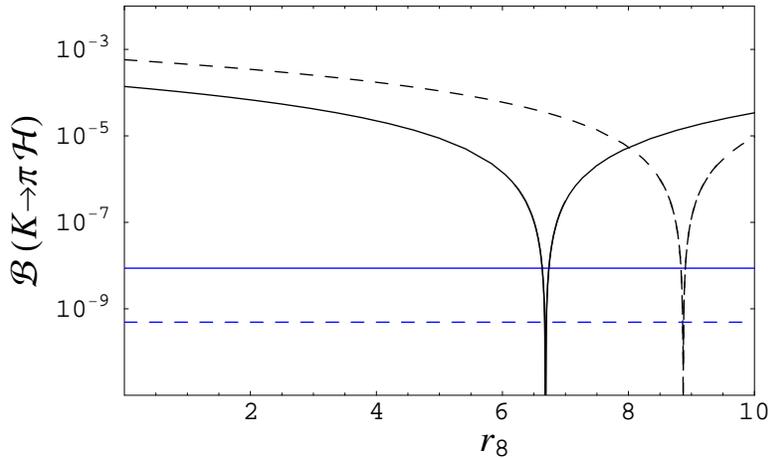} \vspace{-1ex}
\caption{\label{br(KpiH)2hdm}%
Contributions of real parts of total amplitudes for  $\,K^+\to\pi^+\cal H$\, (solid curve)
and  $\,K_L\to\pi^0\cal H$\, (dashed curve) in the 2HDM~II
to their branching ratios as functions of
\,$r_8^{}=\tilde{\gamma}_8^{}/\gamma_8^{}$\,  for  $\,m_{\cal H}^{}=214.3$\,MeV\,
and the parameters in Eq.~(\ref{2hdmpar}).
The horizontal lines indicate the upper bounds in Eqs.~(\ref{K-bound}) and~(\ref{KLbound}).
}
\end{figure}

\begin{figure}[ht]
\vspace{4ex}
\includegraphics[width=4.5in]{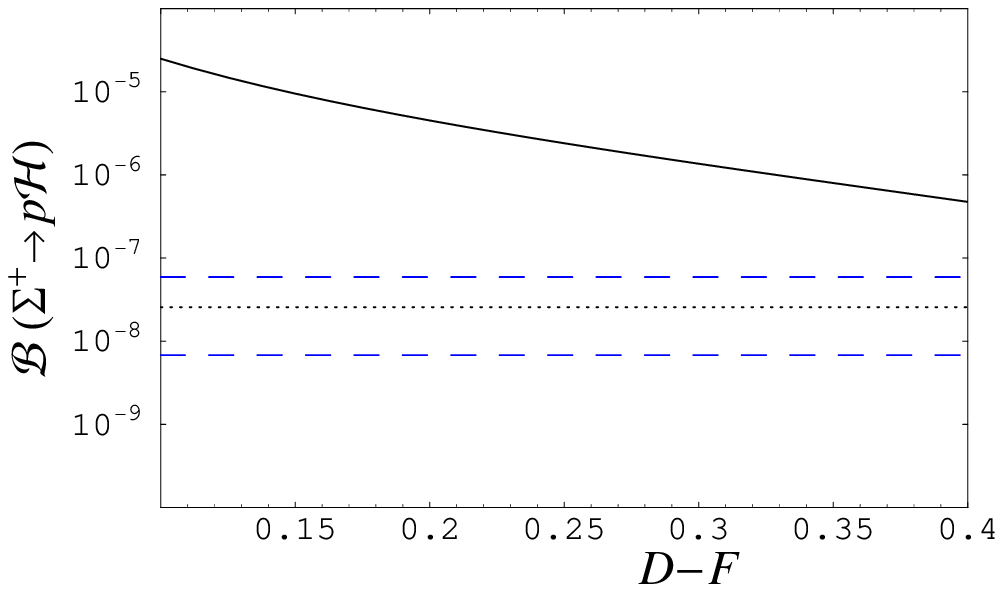} \vspace{-1ex}
\caption{\label{br(SpH)2hdm}%
Contribution of real part of total amplitude for  $\,\Sigma^+\to p\cal H$\, to its branching
ratio in the 2HDM as function of  \,$D$$-$$F$\,  for  $\,m_{\cal H}^{}=214.3$\,MeV\,
and the parameters in Eq.~(\ref{2hdmpar}).
The solid (dotted) curve corresponds to  \,$h_D$$-$$h_F$\, extracted from the
P-wave (S-wave) fit to the  $\,\Sigma^+\to p\pi^0$\,  data using Eqs.~(\ref{ABt})
and~(\ref{ABx}).
The dashed lines correspond to the upper and lower bounds from the HyperCP result.
}
\end{figure}

In Figs.~\ref{br(KpiH)2hdm} and~\ref{br(SpH)2hdm},  we display the corresponding branching
ratios in the 2HDM~II obtained using the parameters in Eqs.~(\ref{gH2hdm}) and~(\ref{2hdmpar}).
In contrast to the SM case, here  \,${\cal B}(K^+\to\pi^+{\cal H})=0$\,  when
\,$r_8^{}\simeq6.7$,\,  but the vanishing of the $K_L$ rate occurs at a different $r_8^{}$
value due to ${\cal M}_{4q}(K_L\to\pi^0{\cal H})$  and
$-{\rm Re}\,{\cal M}_{4q}(K^+\to\pi^+{\cal H})$  being unequal with  \,$k_u\neq k_d$\,
in Eq.~(\ref{kII}).\footnote{We note that, although  $\tilde{\gamma}_8^{}$  is not known from
experiment, there are model calculations~\cite{Leutwyler:1989xj,Bijnens:1998ee} of it yielding
\,$|\tilde{\gamma}_8^{}/\gamma_8^{}|\sim0.2\,$.\,
This would make the kaon rates greatly exceed their bounds, as can be seen from
Figs.~\ref{br(K->piH)} and~\ref{br(KpiH)2hdm}.}
As a consequence, the two kaon constraints cannot be satisfied simultaneously.
Furthermore, the \,$\Sigma^+\to p\cal H$\,  curve that falls  within the HyperCP limits is
the one resulting from the S-wave fit using Eqs.~(\ref{ABt}) and~(\ref{ABx}).

To summarize this section, a light Higgs boson in the SM can be made compatible with
the empirical bounds for \,$\Sigma^+\to p{\cal H}$,\,  while satisfying
the constraints from  \,$K\to\pi{\cal H}$,\,  if the real part of the 2-quark (penguin)
contribution  to the respective amplitudes  is combined with the 4-quark contribution.
Moreover, in the 2HDM such a particle can satisfy all these constraints if its diagonal
couplings to the up- and down-type quarks are the same.
For this to happen in either model, it is necessary for the two amplitudes to cancel
precisely, and we have shown that this is possible for certain values of the hadronic
constants $\tilde{\gamma}_8^{}$,  $h_D$$-$$h_F$, and $D$$-$$F$.
Although $\tilde{\gamma}_8^{}$ is not known, unlike $h_D$$-$$h_F$ and $D$$-$$F$
which are extractable from hyperon nonleptonic and semileptonic decays~\cite{bcpt},
it has a definite value in the SM and cannot be fine-tuned.
We note that in all the \,$\Sigma^+\to p\cal H$\, cases discussed above
the  $\bar{p}\gamma_5^{}\Sigma^+$  term in the amplitude is small compared to
the $\bar{p}\Sigma^+$ term and that, therefore, the $\tilde{\gamma}_8^{}$
contributions are important only in the kaon cases.
We also note that flipping the signs of $A_{p\pi^0}$ and $B_{p\pi^0}$ in Eq.~(\ref{ABx}),
whose overall sign is not fixed by experiment, would prevent the cancelation in the
hyperon case from occurring and thus resulting in rates much above the bounds.

It turns out that the imaginary part of the penguin amplitude is sufficient to eliminate
these scalar particles as candidates for the HyperCP events, as it cannot be canceled by
the 4-quark amplitudes~\cite{Cheng:1989ib}, having a size of
\begin{eqnarray}
|{\rm Im}\,g_{\cal H}^{}| &\sim& 5.8 \times 10^{-7}  \,\,,
\end{eqnarray}
much larger than allowed by Eq.~(\ref{kbound}) with  \,${\rm Im}\,M_{4K}=0$.\,
The scaling of the penguin amplitude to the $B$-meson system is also incompatible with
the  \,$b\to s X$\,  bound.
In the SM
\begin{eqnarray}
g_{\cal H}^\prime &=& \frac{3\,G_F^{}}{16\sqrt2\,\pi^2}\sum_{q=u,c,t}m_q^2 V^*_{qs}V_{qb}^{}
\,\,\sim\,\,   -1.7 \times 10^{-4}  \,\,,
\end{eqnarray}
which is much larger than allowed by  Eq.~(\ref{ggbs}).
In the 2HDM, the relative size is also too large:
\,$|g_{\cal H}^\prime/g_{\cal H}^{}|\sim |V_{tb}/V_{td}|$.\footnote{One could arrive at
a similar conclusion about $\cal H$ in the 2HDM~II by analyzing the decay
\,$\eta\to\pi^0\cal H$,\, whose amplitude depends on the 4-quark parameters
$\,k_d^{}$$-$$k_u^{}$~\cite{Prades:1990vn}.  Thus, from the 90\%-C.L. bound
\,${\cal B}(\eta\to\pi^0{\cal H})<5\times10^{-6}$~\cite{Dzhelyadin:1980ti}, one extracts
\,$|k_d^{}-k_u^{}|<0.45$\,  for  \,$m_{\cal H}^{}=214.3$\,MeV,\,  which is incompatible with
the limit derived from  Eq.~(\ref{kII}) plus the LEP constraint
\,$\sin^2(\beta-\alpha)<0.06$~\cite{Abbiendi:2004gn}, namely
\,$|k_d^{}-k_u^{}|=|2\,\cos(\alpha-\beta)/\sin(2\beta)|>1.9.$}

Both of these problems are associated with a structure in which the Higgs-penguin amplitude
is dominated by diagrams with up-type quarks and $W$ bosons in the loops.
It may be possible to remedy these problems in models with additional contributions to
the penguin, for example, from supersymmetric (SUSY) partners.
If the penguin can be sufficiently suppressed, Eqs.~(\ref{MSpH'}) and~(\ref{M4q[K->pi-H]})
suggest that models in which  \,$k_W\sim k_G$\,  could satisfy the kaon bounds while being able
to account for the HyperCP result.

\section{Pseudoscalar Higgs boson\label{pseudoscalarH}}

We now consider the possibility that the new particle is a light $CP$-odd pseudoscalar,
$\cal A$, in the two-Higgs-doublet model.
Specifically, we do so in types~I and~II of the model.

\subsection{Two-quark $\bm{|\Delta S|=1}$  interactions}

The 2-quark flavor-changing couplings of  ${\cal A}$ in the 2HDM are induced at one loop
and have been evaluated in Refs.~\cite{Frere:1981cc,Hall:1981bc}.
The effective Lagrangian is the same in types I and II of the model and can be written as
${\cal L}_{{\cal A}sd}$  in Eq.~(\ref{Asd}), where
\begin{eqnarray}   \label{g_P}
g_{\cal A}^{} \,\,=\,\,
\frac{G_F^{}}{16\sqrt2\,\pi^2}\sum_{q=u,c,t}^{}m_q^2\,V_{qd}^*V_{qs}^{}\,
\Biggl(\frac{A_1^{}(q)}{\tan\beta}+\frac{A_2^{}(q)}{\tan^3\beta}\Biggr)  \,\,,
\end{eqnarray}
with  $A_{1,2}(q)$  being functions of $m_q^{}$, $m_W^{}$, and  $m_{H^+}$,
whose expressions can be found in Ref.~\cite{Frere:1981cc}.
The leading-order chiral realization of  ${\cal L}_{{\cal A}sd}$ is then
\begin{eqnarray}   \label{LsdP}
{\cal L}_{\cal A}^{}  &=&
b_D^{} \left\langle \bar{B}{}^{} \left\{ h_{\cal A}^{}, B^{} \right\} \right\rangle
+ b_F^{} \left\langle \bar{B}{}^{} \left[ h_{\cal A}^{}, B^{} \right] \right\rangle
+ b_0^{} \left\langle h_{\cal A}^{} \right\rangle \left\langle\bar{B}{}^{}B^{}\right\rangle
\,+\,  \mbox{$\frac{1}{2}$} f^2 B_0^{} \left\langle h_{\cal A}^{} \right\rangle
\,\,+\,\,  {\rm H.c.}    \,\,,
\end{eqnarray}
where
\begin{eqnarray}
h_{\cal A}^{}  \,\,=\,\,
-2{i}g_{\cal A}^{}\,\bigl(\xi^\dagger h M\xi^\dagger-\xi M h\xi\bigr)\frac{\cal A}{v} \,\,.
\end{eqnarray}

The leading-order diagrams for  $\,K\to\pi\cal A$\,  and   $\,\Sigma^+\to p\cal A$\,
arising from  ${\cal L}_{\cal A}$, plus ${\cal L}_{\rm s}$,  are similar to those in the case
of standard-model Higgs boson, displayed in Fig.~\ref{sdHdiagrams}.
The resulting amplitudes are
\begin{eqnarray}   \label{MKpiP}
{\cal M}_{2q}^{}(K^+\to\pi^+{\cal A})  \,\,=\,\,
-\sqrt2\,{\cal M}_{2q}^{}(K^0\to\pi^0{\cal A})
\,\,=\,\, {i}g_{\cal A}^{}\,\frac{m_K^2-m_\pi^2}{v}  \,\,,
\end{eqnarray}
\begin{eqnarray}   \label{MSpP}
{\cal M}_{2q}^{}(\Sigma^+\to p {\cal A})  \,\,=\,\,
{i}g_{\cal A}^{}\,\frac{m_\Sigma^{}-m_N^{}}{v}\, \bar{p}\Sigma^+
\,-\,
{i}g_{\cal A}^{}\,(D-F)\,\frac{m_\Sigma^{}+m_N^{}}{v}\,\frac{m_K^2}{m_K^2-m_{\cal A}^2}\,
\bar{p}\gamma_5^{}\Sigma^+  \,\,.
\end{eqnarray}

\subsection{Four-quark $\bm{|\Delta S|=1}$  interactions}

The diagonal couplings of ${\cal A}$ to light quarks in the 2HDM  are described
by~\cite{Hall:1981bc}
\begin{eqnarray}  \label{L_qqP}
{\cal L}_{{\cal A}qq} \,\,=\,\,  -\bar{q}\tilde{M}\gamma_5^{}q\, \frac{{i}{\cal A}}{v}
\,\,=\,\,
-\bar{q}_{\rm L}^{}\tilde{M}q_{\rm R}^{}\,\frac{{i}{\cal A}}{v} \,\,+\,\, {\rm H.c.}  \,\,,
\end{eqnarray}
where
\begin{eqnarray}
q  \,\,=\,\, (u\,\,\,d\,\,\,s)^{\rm T}   \,\,,  \hspace{2em}
\tilde{M}  \,\,=\,\, {\rm diag}\bigl(l_u^{}{}\hat{m},\,l_d^{}{}\hat{m},\,l_d^{}{}m_s^{}\bigr) \,\,,
\end{eqnarray}
with
\begin{eqnarray}  \label{lI}
l_u^{}{}  \,\,=\,\, -l_d^{}{}  &=&  -\cot\beta  \hspace{3em}  \mbox{in the 2HDM~I} \,\,,
\\  \label{lII}
l_u^{}{}  \,\,=\,\, -\cot\beta  \,\,, &&
l_d^{}{}  \,\,=\,\, -\tan\beta  \hspace{3em}  \mbox{in the 2HDM~II} \,\,.
\end{eqnarray}
Since the Lagrangian for the quark masses is
\,${\cal L}_q=-\bar{q}_{\rm L}^{}Mq_{\rm R}^{}\,+\, {\rm H.c.}$,\,
the effect of  ${\cal L}_{{\cal A}qq}$  on interactions described by  ${\cal L}_{\rm s,w}$
can be taken into account using  ${\cal L}_{\rm s,w}$  and substituting  $M$  with
\,$\tilde{M}{i}{\cal A}/v$\,~\cite{Grzadkowski:1992av}.
The resulting Lagrangians are
\begin{eqnarray}   \label{LsP}
{\cal L}_{\rm s}^{\cal A}  \,\,=\,\,
\left( b_D^{}\,\bigl\langle \bar{B}^{} \bigl\{ \tilde{M}_-^{}, B^{} \bigr\} \bigr\rangle
+ b_F^{}\, \bigl\langle \bar{B}^{} \bigl[ \tilde{M}_-^{}, B^{} \bigr] \bigr\rangle
+ b_0^{}\, \bigl\langle \tilde{M}_-^{} \bigr\rangle \bigl\langle \bar{B}^{} B^{} \bigr\rangle
\,+\, \mbox{$\frac{1}{2}$} f^2 B_0^{}\, \bigl\langle\tilde{M}_-^{}\bigr\rangle \right)
\frac{{i}{\cal A}}{v}  \,\,,
\end{eqnarray}
\begin{eqnarray} \label{LwP}
{\cal L}_{\rm w}^{\cal A}  \,\,=\,\,
2\tilde{\gamma}_8^{}\, f^2 B_0^{}\, \bigl\langle h \xi \tilde{M}_-^{}\xi^\dagger
\bigr\rangle \frac{{i}{\cal A}}{v}
\,\,+\,\,  {\rm H.c.}   \,\,,
\end{eqnarray}
where
\begin{eqnarray}
\tilde{M}_-^{} \,\,=\,\, \xi^\dagger\tilde{M}\xi^\dagger-\xi\tilde{M}^\dagger\xi  \,\,.
\end{eqnarray}
In addition, if the SU(3) singlet $\eta_1^{}$ is included in  ${\cal L}_{\rm s,w}^{({\cal A})}$ by
replacing  $\Sigma$ with  \,$\Sigma\,\exp\bigl(i\sqrt{2/3}\,\eta_1^{}/f\bigr)$,\,
the coupling of  ${\cal A}$  to two gluons via the axial anomaly gives rise
to~\cite{Grzadkowski:1992av}
\begin{eqnarray} \label{LeP}
{\cal L}_{\eta_1^{}{\cal A}}^{}  \,\,=\,\,
-\mbox{$\frac{1}{2}$}
\Bigl(m_{\eta_1^{}}^2-\mbox{$\frac{2}{3}$}m_K^2-\mbox{$\frac{1}{3}$}m_\pi^2\Bigr)
\Biggl[\eta_1^{}+\frac{f\,{\cal A}}{\sqrt6\,v}(2l_u^{}{}+l_d^{}{})\Biggr]^2  \,\,,
\end{eqnarray}
which modifies the $\eta_1^{}$-${\cal A}$ mixing generated by  ${\cal L}_{\rm s}^{\cal A}$.

\begin{figure}[b]
\begin{picture}(120,90)(-60,-10)
\Text(-32,0)[r]{\footnotesize$K$} \Line(-30,0)(30,0)
\Line(0,0)(0,30) \Text(0,35)[]{\footnotesize${\cal A}$}
\Text(32,0)[l]{\footnotesize$\pi$} \CBoxc(0,0)(4,4){Black}{Black}
\end{picture}
\begin{picture}(120,90)(-60,-10)
\Text(-32,0)[r]{\footnotesize$K$} \Line(-30,0)(30,0) \Line(0,0)(20,20) \Vertex(0,0){2}
\Line(0,0)(0,30) \Text(0,35)[]{\footnotesize${\cal A}$} \Text(19,9)[]{\footnotesize$K^0$}
\Text(32,0)[l]{\footnotesize$\pi$} \CBoxc(20,20)(4,4){Black}{Black}
\end{picture}
\begin{picture}(120,90)(-60,-10)
\Text(-32,0)[r]{\footnotesize$K$} \Line(-30,0)(30,0) \Line(0,0)(20,20) \Vertex(0,30){2}
\Line(0,0)(0,50) \Text(0,55)[]{\footnotesize${\cal A}$} \Text(19,9)[]{\footnotesize$K$}
\Text(-2,15)[r]{\footnotesize$\pi^0,\eta,\eta'$} \Text(32,0)[l]{\footnotesize$\pi$}
\CBoxc(20,20)(4,4){Black}{Black} \Vertex(0,0){2}
\end{picture}
\begin{picture}(120,90)(-60,-10)
\Text(-32,0)[r]{\footnotesize$K$} \Line(-30,0)(30,0) \Vertex(0,30){2}
\Line(0,0)(0,50) \Text(0,55)[]{\footnotesize${\cal A}$} \CBoxc(0,0)(4,4){Black}{Black}
\Text(2,15)[l]{\footnotesize$\pi^0,\eta,\eta'$} \Text(32,0)[l]{\footnotesize$\pi$}
\end{picture}   \vspace*{-1ex}
\caption{\label{4qPkaon}
Diagrams contributing to  $\,K\to\pi {\cal A}$\,  arising from  the 4-quark operators.
The dots come from  ${\cal L}_{\rm s}^{({\cal A})}$  in  Eqs.~(\ref{Lstrong}) and~(\ref{LsP}),
whereas the square vertices are from  ${\cal L}_{\rm w}^{({\cal A})}$  in  Eqs.~(\ref{Lweak})
and~(\ref{LwP}).
}
\end{figure}
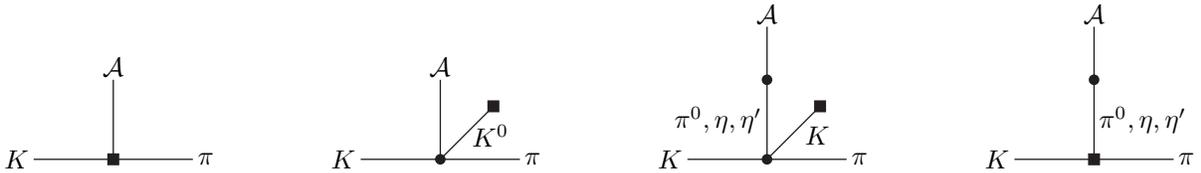

 {From}  ${\cal L}_{\rm s,w}^{({\cal A})}$, we derive the leading-order
diagrams shown in Fig.~\ref{4qPkaon} for \,$K\to\pi {\cal A}$,\,  where
\begin{eqnarray}
\eta  \,\,=\,\,  \eta_8^{}\, \cos\theta - \eta_1^{}\,\sin\theta   \,\,,  \hspace{2em}
\eta'  \,\,=\,\,  \eta_8^{}\,\sin\theta + \eta_1^{}\, \cos\theta   \,\,.
\end{eqnarray}
The resulting amplitudes are
\begin{subequations}  \label{MKpiP'}
\begin{eqnarray}
&& \hspace*{-5em} {\cal M}_{4q}^{}\bigl(K^+\to\pi^+ {\cal A}\bigr)  \,\,=\,\,
\frac{i\gamma_8^{}\, (l_u^{}{}-l_d^{}{})\,m_\pi^2}{2v}
\nonumber \\ && +\,\,
i\gamma_8^{}\,\bigl[\bigl(2m_K^2+m_\pi^2-3m_{\cal A}^2\bigr)\,c_\theta^{}
- \sqrt8\,\bigl(m_K^2-m_\pi^2\bigr)\,s_\theta^{}\bigr]
\nonumber \\ && \,\,\,\, \times\,\,
\frac{\bigl[4l_d^{}{}\,m_K^2-\bigl(3l_d^{}{}+l_u^{}{}\bigr)\,m_\pi^2\bigr]\,c_\theta^{}
      + \sqrt2\,\bigl[2l_d^{}{}\, m_K^2 + l_u^{}{}\,m_\pi^2
                      - \bigl(l_d^{}{}+2l_u^{}{}\bigr)\,\tilde{m}_0^2\bigr]\,s_\theta^{}}
{6\bigl(m_\eta^2-m_{\cal A}^2\bigr)\,v}
\nonumber \\ && +\,\,
i\gamma_8^{}\,\bigl[\bigl(2m_K^2+m_\pi^2-3m_{\cal A}^2\bigr)\,s_\theta^{}
+\sqrt8\,\bigl(m_K^2-m_\pi^2\bigr)\,c_\theta^{}\bigr]
\nonumber \\ && \,\,\,\, \times\,\,
\frac{\bigl[4l_d^{}{}\,m_K^2-\bigl(3l_d^{}{}+l_u^{}{}\bigr)\,m_\pi^2\bigr]\,s_\theta^{}
      - \sqrt2\, \bigl[2l_d^{}{}\,m_K^2+l_u^{}{}\,m_\pi^2
                       - \bigl(l_d^{}{}+2l_u^{}{}\bigr)\,\tilde{m}_0^2\bigr]\, c_\theta^{}}
{6\bigl(m_{\eta'}^2-m_{\cal A}^2\bigr)\,v}  \,\,,
\end{eqnarray}
\begin{eqnarray}
&& \hspace*{-5em} {\cal M}_{4q}^{}\bigl(K^0\to\pi^0{\cal A}\bigr)  \,\,=\,\,
\frac{i\gamma_8^{}\,\bigl(l_u^{}{}-l_d^{}{}\bigr)\,\bigl(2m_K^2-m_\pi^2-m_{\cal A}^2\bigr)\,m_\pi^2}
     {\sqrt8\,\bigl(m_{\cal A}^2-m_\pi^2\bigr)\,v}
\nonumber \\ && +\,\,
i\gamma_8^{}\,\bigl[ \bigl(2m_K^2+m_\pi^2-3m_{\cal A}^2\bigr)\,c_\theta^{}
                    - \sqrt8\, \bigl(m_K^2-m_\pi^2\bigr)\, s_\theta^{} \bigr]
\nonumber \\ && \,\,\,\, \times\,\,
\frac{\bigl[4l_d^{}{}\,m_K^2-\bigl(3l_d^{}{}+l_u^{}{}\bigr)\,m_\pi^2\bigr]\,c_\theta^{}
+ \sqrt2\,\bigl[2l_d^{}{}\,m_K^2+l_u^{}{}\,m_\pi^2-\bigl(l_d^{}{}+2l_u^{}{}\bigr)\,\tilde{m}_0^2\bigr]\,s_\theta^{}}
{6\sqrt2\,\bigl(m_{\cal A}^2-m_\eta^2\bigr)\,v}
\nonumber \\ && +\,\,
i\gamma_8^{}\,\bigl[ \bigl(2m_K^2+m_\pi^2-3m_{\cal A}^2\bigr)\, s_\theta^{}
                    + \sqrt8\, \bigl(m_K^2-m_\pi^2\bigr)\, c_\theta^{} \bigr]
\nonumber \\ && \,\,\,\, \times\,\,
\frac{\bigl[4l_d^{}{}\,m_K^2-\bigl(3l_d^{}{}+l_u^{}{}\bigr)\,m_\pi^2\bigr]\,s_\theta^{}
- \sqrt2\,\bigl[2l_d^{}{}\,m_K^2+l_u^{}{}\,m_\pi^2-\bigl(l_d^{}{}+2l_u^{}{}\bigr)\,\tilde{m}_0^2\bigr]\,c_\theta^{}}
{6\sqrt2\,\bigl(m_{\cal A}^2-m_{\eta'}^2\bigr)\,v}  \,\,,
\end{eqnarray}
\end{subequations}
where
\begin{eqnarray}
c_\theta^{}  \,\,=\,\, \cos\theta  \,\,,  \hspace{2em}
s_\theta^{}  \,\,=\,\, \sin\theta  \,\,,  \hspace{2em}
\tilde{m}_0^2  \,\,=\,\,  m_{\eta_1}^2-\mbox{$\frac{2}{3}$}m_K^2-\mbox{$\frac{1}{3}$}m_\pi^2 \,\,.
\end{eqnarray}
The $\tilde{\gamma}_8^{}$ contributions to this amplitude cancel
completely, as already noted in Ref.~\cite{Grzadkowski:1992av}.
Numerically,  $\,\tilde{m}_0^{}\simeq819$\,MeV\, from fitting to the $\eta'$ mass after
diagonalizing the $\eta_{8,1}^{}$ masses derived from the Lagrangians in Eqs.~(\ref{Lstrong})
and~(\ref{LeP}), and consequently  $\,\theta\simeq-19.7^\circ$.\,

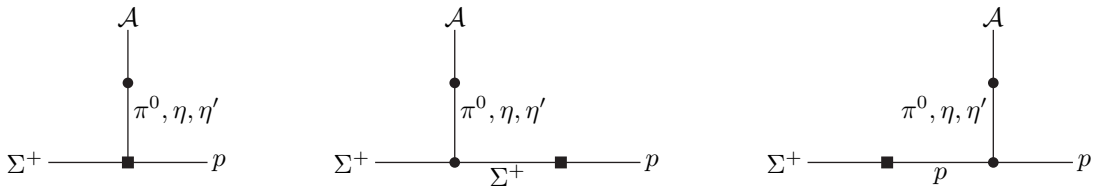
\begin{figure}[b]
\begin{picture}(120,100)(-60,-20)
\Text(-32,0)[r]{\footnotesize$\Sigma^+$} \Line(-30,0)(30,0)
\Line(0,0)(0,50) \Text(0,55)[]{\footnotesize${\cal A}$} \Vertex(0,30){2}
\Text(32,0)[l]{\footnotesize$p$} \CBoxc(0,0)(4,4){Black}{Black}
\Text(2,20)[l]{\footnotesize$\pi^0,\eta,\eta'$}
\end{picture}
\begin{picture}(160,100)(-80,-20)
\Text(-52,0)[r]{\footnotesize$\Sigma^+$} \Line(-50,0)(50,0)
\Line(-20,0)(-20,50) \Text(-20,55)[]{\footnotesize${\cal A}$}
\Text(0,-5)[]{\footnotesize$\Sigma^+$} \Vertex(-20,0){2} \Vertex(-20,30){2}
\Text(52,0)[l]{\footnotesize$p$} \CBoxc(20,0)(4,4){Black}{Black}
\Text(-18,20)[l]{\footnotesize$\pi^0,\eta,\eta'$}
\end{picture}
\begin{picture}(160,100)(-80,-20)
\Text(-52,0)[r]{\footnotesize$\Sigma^+$} \Line(-50,0)(50,0)
\Line(20,0)(20,50) \Text(20,55)[]{\footnotesize${\cal A}$}
\Text(0,-5)[]{\footnotesize$p$} \Vertex(20,0){2} \Vertex(20,30){2}
\Text(52,0)[l]{\footnotesize$p$} \CBoxc(-20,0)(4,4){Black}{Black}
\Text(18,20)[r]{\footnotesize$\pi^0,\eta,\eta'$}
\end{picture}   \vspace*{-2ex}
\caption{\label{4qPdiagrams'}%
Diagrams contributing to  $\,\Sigma^+\to p\cal A$\,  arising from  the 4-quark operators.
The square vertices come from  ${\cal L}_{\rm w}$  in  Eq.~(\ref{Lweak}), whereas the dots
are from the Lagrangians in  Eqs.~(\ref{Lstrong}), (\ref{LsP}), and~(\ref{LeP}).
}
\end{figure}

The leading-order 4-quark contributions to  \,$\Sigma^+\to p\cal A$\,  arise from the diagrams
in  Fig.~\ref{4qPdiagrams'} and can be expressed as
\begin{eqnarray}   \label{MSpP'}
{\cal M}_{4q}^{}(\Sigma^+\to p {\cal A})  &=&
{i}\bar{p}\, \bigl(A_{p{\cal A}}^{}-B_{p{\cal A}}^{}\gamma_5^{}\bigr)\,\Sigma^+   \;,
\end{eqnarray}
where
\begin{eqnarray}
A_{p{\cal A}}^{}  &=&
\frac{f\,A_{p\pi^0}^{}\,\bigl(l_d^{}{}-l_u^{}{}\bigr)\,m_\pi^2}
     {2\bigl(m_{\cal A}^2-m_\pi^2\bigr)\,v}
\nonumber \\ && \!\!\! +\,\,
\frac{f\,A_{p\pi^0}^{} \left\{
\bigl[4l_d^{}{}\,m_K^2-\bigl(3l_d^{}{}+l_u^{}{}\bigr)\,m_\pi^2\bigr]\,c_\theta^2
+ \sqrt2\, \bigl[ 2l_d^{}{}\,m_K^2+l_u^{}{}\,m_\pi^2-\bigl(l_d^{}{}+2l_u^{}{}\bigr)\,\tilde{m}_0^2\bigr]\,
c_\theta^{}s_\theta^{} \right\}}
{2 \bigl(m_\eta^2-m_{\cal A}^2\bigr)\, v}
\nonumber \\ && \!\!\! +\,\,
\frac{f\,A_{p\pi^0}^{} \left\{
\bigl[4l_d^{}{}\,m_K^2-\bigl(3l_d^{}{}+l_u^{}{}\bigr)\,m_\pi^2\bigr]\,s_\theta^2
- \sqrt2\,\bigl[2l_d^{}{}\,m_K^2+l_u^{}{}\,m_\pi^2-\bigl(l_d^{}{}+2l_u^{}{}\bigr)\,\tilde{m}_0^2\bigr]\,
c_\theta^{}s_\theta^{} \right\}}
{2 \bigl(m_{\eta'}^2-m_{\cal A}^2\bigr)\, v}  \,\,,
\hspace*{2em}
\end{eqnarray}
\begin{eqnarray}
B_{p{\cal A}}^{}  &=&
\frac{f\,B_{p\pi^0}^{}\,\bigl(l_d^{}-l_u^{}\bigr)\,m_\pi^2}{2\bigl(m_{\cal A}^2-m_\pi^2\bigr)\,v}
\nonumber \\ && \!\!\! +\,\,
\frac{f\,B_{p\pi^0}^{} \left\{
\bigl[4l_d^{}\,m_K^2-\bigl(3 l_d^{}+l_u^{}\bigr)\,m_\pi^2\bigr]\,c_\theta^2
+ \sqrt2\, \bigl[ 2l_d^{}\,m_K^2+l_u^{}\,m_\pi^2-\bigl(l_d^{}+2l_u^{}\bigr)\,\tilde{m}_0^2\bigr]\,
c_\theta^{}s_\theta^{} \right\}}
{2 \bigl(m_\eta^2-m_{\cal A}^2\bigr)\, v}
\nonumber \\ && \!\!\! +\,\,
\frac{f\,B_{p\pi^0}^{} \left\{
\bigl[4l_d^{}\,m_K^2-\bigl(3 l_d^{}+l_u^{}\bigr)\,m_\pi^2\bigr]\,s_\theta^2
- \sqrt2\, \bigl[ 2l_d^{}\,m_K^2+l_u^{}\,m_\pi^2-\bigl(l_d^{}+2l_u^{}\bigr)\,\tilde{m}_0^2\bigr]\,
c_\theta^{}s_\theta^{} \right\}}
{2 \bigl(m_{\eta'}^2-m_{\cal A}^2\bigr)\, v}  \,\,,
\hspace*{2em}
\end{eqnarray}
where  $A_{p\pi^0}$  and  $B_{p\pi^0}$  are given in Eq.~(\ref{ABt}).
We note that contributions with  $\gamma_8^{}$ or  $\tilde{\gamma}_8^{}$  appear only
at next-to-leading order.

\subsection{Total contributions}

\begin{figure}[t]
\vspace{2ex} \hspace*{-2em}
\includegraphics[width=7in]{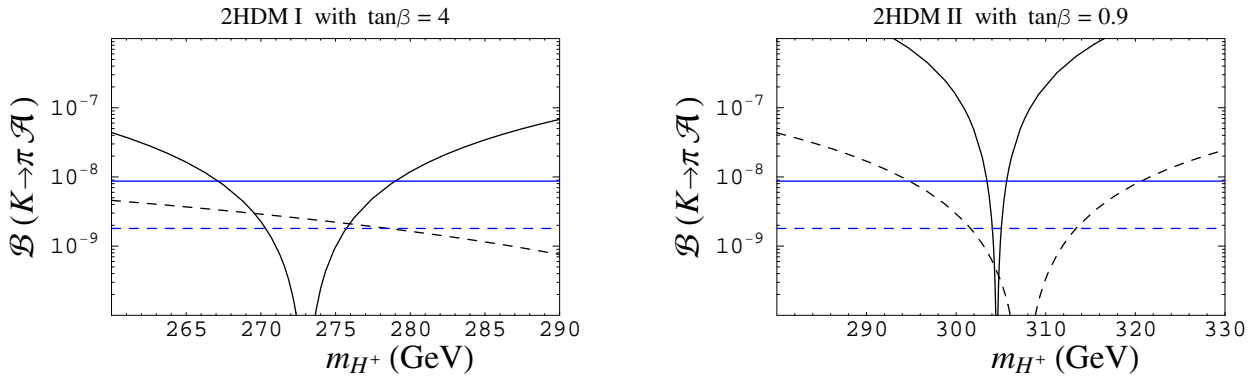}   \vspace*{-1em}
\caption{\label{br(K->piP)}%
Contributions of real parts of total amplitudes for  $\,K^+\to\pi^+\cal A$\, (solid curve)
and  $\,K_S\to\pi^0\cal A$\, (dashed curve) in the 2HDM to their branching ratios as
functions of charged-Higgs-boson mass for  $\,m_{\cal A}^{}=214.3$\,MeV\,
and $\,\tan\beta=4\,(0.9)$ in type I\,(II) of the model.
The horizontal lines indicate the upper bounds in Eqs.~(\ref{K-bound}) and~(\ref{KSbound}).
}
\end{figure}

\begin{figure}[ht] \vspace{1ex}
\includegraphics[width=4in]{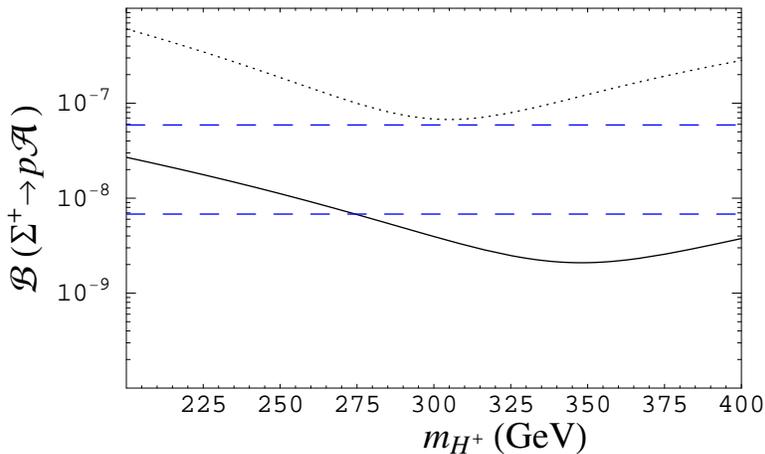}   \vspace*{-1em}
\caption{\label{br(S->pP)}%
Contribution of real part of total amplitude for $\,\Sigma^+\to p\cal A$\, to its branching
ratio in the 2HDM~I (solid curve) and~II~(dotted curve) as function of charged-Higgs-boson
mass for  $\,m_{\cal A}^{}=214.3$\,MeV\, and $\,\tan\beta=4\,(0.9)$ in type I\,(II).
The dashed lines indicate the bounds from the HyperCP result.
}
\end{figure}

The total amplitudes for  \,$K\to\pi\cal A$\, result from adding the contributions
in Eqs.~(\ref{MKpiP}) and~(\ref{MKpiP'}).
If the $CP$-violating terms in the amplitudes are ignored, it is possible for
the 2-quark and 4-quark contributions to cancel.
We show this possibility in Fig.~\ref{br(K->piP)}, where we plot the resulting branching ratios
as functions of the charged-Higgs-boson mass for  $\,m_{\cal A}^{}=214.3$\,MeV\,  and  different
$\tan\beta$ values in the 2 versions of the 2HDM.
The total amplitude for  \,$\Sigma^+\to p {\cal A}$\, is the sum of the contributions in
Eqs.~(\ref{MSpP}) and~(\ref{MSpP'}).
If the experimental values of  $A_{p\pi^0}$  and  $B_{p\pi^0}$  in Eq.~(\ref{ABx}),
as well as the middle value  $\,D-F=0.25$,\, are used in the total amplitude,
the resulting branching ratios in the 2HDM are displayed in Fig.~\ref{br(S->pP)}.


We have found that only one of the kaon bounds can be satisfied if the HyperCP
result is assumed to be mediated by $\cal A$ in the 2HDM.
However, for certain  $\tan\beta$ and $m_{H^+}$  values near the ones indicated in
Figs.~\ref{br(K->piP)} and~\ref{br(S->pP)}, all the kaon and hyperon constraints
can be nearly satisfied simultaneously.
Part of the difficulty in satisfying all of the constraints lies with the vanishing of
the $K^+$ and $K_S$ rates occurring at different $m_{H^+}$ values, which is due to
${\cal M}_{4q}(K_S\to\pi^0{\cal A})$ and  $-{\rm Re}\,{\cal M}_{4q}(K^+\to\pi^+{\cal A})$
being unequal with  $\,l_u\neq l_d$\,  in Eqs.~(\ref{lI}) and~(\ref{lII}).
We note that the situation is not much different if the  signs of $A_{p\pi^0}$ and
$B_{p\pi^0}$ in Eq.~(\ref{ABx}) are both flipped.

To summarize this section, we have found that it is possible for the real part of
the penguin amplitude to cancel against the 4-quark amplitude to approximately satisfy
the kaon bounds while explaining the HyperCP observation with a 2HDM pseudoscalar.
Unlike the scalar case, there is no free hadronic parameter at leading order in $\chi$PT
in this case.
The cancelation must happen as a function of the short-distance parameters that determine
the size of the amplitudes.

A feature shared by scalars and pseudoscalars in the 2HDM is that the imaginary part of
the penguin amplitude is incompatible with the kaon bounds in Eq.~(\ref{kbound}) and has no
counterpart that could cancel it in the 4-quark amplitude.
A related problem is that the scaling of the penguin amplitude to the $B$ system is also
incompatible with observation.

In view of these flaws, it is tempting to search for a model in which the penguin
amplitudes are completely suppressed, and the 2HDM~II seems to allow us to do that.
In the 2HDM II the penguin amplitudes are proportional to $l_u$, whereas the 4-quark
amplitudes receive contributions from both $l_u$ and $l_d$ in Eq.~(\ref{lII}).
Thus the model in the large-$\tan\beta$ limit has  \,$l_u\to 0$.\,
Unfortunately, in this limit $l_d$ induces 4-quark amplitudes resulting in
\begin{eqnarray}
\frac{{\cal B}(\Sigma^+ \to p{\cal A})}{{\cal B}(K^+\to\pi^+{\cal A})}  &\to& 0.025  \,\,,
\end{eqnarray}
which is inconsistent with Eqs.~(\ref{kbound}) and~(\ref{sigbound}).
In the 2HDM~I, which has $l_u$ and $l_d$ given in Eq.~(\ref{lI}), the 4-quark
amplitudes alone yield
\begin{eqnarray}
\frac{{\cal B}_{4q}^{}(\Sigma^+\to p{\cal A})}{{\cal B}_{4q}^{}(K^+\to\pi^+{\cal A})} &=& 0.53
\end{eqnarray}
for all values of $\tan\beta$, which is consistent with Eqs.~(\ref{kbound}) and~(\ref{sigbound}).
However, in this case it is the penguin amplitude that eliminates the pseudoscalar as
a possible HyperCP candidate.

These results suggest the ingredients of a model that can satisfy all constraints.
It is necessary for the penguin amplitudes to be dominated by additional particles, such as
SUSY partners, in such a way that $g_{\cal A}$ is not proportional to top-quark CKM angles.
We have sketched a scenario where this happens in Ref.~\cite{usletter}.

\section{Summary and Conclusions\label{final}}

We have summarized the existing constraints on the production of a light Higgs boson in kaon and
$B$-meson decays, as well as the implication of attributing the HyperCP events to the production
of a light Higgs boson in hyperon decay.

Production rates for such a particle in kaon and hyperon decays receive contributions from two-
and four-quark operators that can be comparable in some cases.
We have investigated the interplay of both production mechanisms with the aid of leading-order
chiral perturbation theory.
To this effect, we have implemented the low-energy theorems governing the couplings of light
(pseudo)scalars to hadrons at leading order in baryon $\chi$PT, generalizing
existing studies for kaon decay.

We first discussed the case of a scalar Higgs boson. We found that the leading-order amplitudes
in both kaon and hyperon decays depend on an unknown low-energy constant $\tilde\gamma_8^{}$,
as well as known constants from the hyperon sector.
This constant is connected to a weak-mass term in the chiral Lagrangian that can be
rotated away for processes that involve only pseudo-Goldstone bosons and is, therefore, unknown.
We applied our results to the process  \,$\Sigma^+\to p X$\, relevant to the HyperCP
observation  of  \,$\Sigma^+\to p\mu^+\mu^-$.\,
We showed that the two-quark contributions in the SM and its 2HDM extensions are too large to
explain the HyperCP observation.
However, we also showed that there can be cancelations between the $CP$-conserving two- and
four-quark contributions to this process that lead to a rate comparable in size to the HyperCP
observation for both the SM and the 2HDM.
Such cancelations occur for a certain range of known constants from the hyperon sector,
the effect of $\tilde\gamma_8^{}$ being small.
In both cases, however, the two-quark penguin contribution has an imaginary
($CP$ violating) part that is too large to be compatible with the HyperCP result.
In the SM and in the 2HDM, the four-quark contributions have a $CP$-violating part that is
much smaller than that of the penguin amplitude and hence these models are ruled out as
explanations for the HyperCP observation.
More general models with additional $CP$-violating phases may be able to address this issue.
In addition, in these models the scaling of the two-quark operator to the $B$ system  is
incompatible with the nonobservation of a light scalar in $B$ decay.

We then discussed the case of a pseudoscalar Higgs boson in the 2HDM.
In this case we computed the leading-order amplitudes in $\chi$PT and included, as well,
certain higher-order terms mediated by the $\eta^\prime$ state.
The resulting amplitudes for both kaon and hyperon decays do not depend on any unknown
hadronic parameters.
In particular, they do not depend on $\tilde\gamma_8^{}$, as observed in
Ref.~\cite{Grzadkowski:1992av}.
We then applied our results to the  \,$\Sigma^+\to p\cal A$\,  process.
Once again we found that the real part of the amplitude can be consistent with the HyperCP
observation for a certain range of parameters in the 2HDM  ($\tan\beta$ and $m_{H^+}$),
but that the imaginary part of the penguin amplitude is too large.
The scaling of the two-quark operator to the $B$ system also produces a $B\to X_s^{}\cal A$
rate that is too large. Both of these problems can be solved in more general models that
modify the phase and scaling with CKM angles of the two-quark operator.

In conclusion, we have shown that it is possible to interpret the HyperCP observation as
evidence for a light Higgs boson, although it is not easy to arrange this in a model.
Typical Higgs-penguin operators have three problems:
\vspace*{-1ex}
\newcounter{num}
\begin{list}{(\alph{num})}{\usecounter{num}}
\item if they have the right size to fit the HyperCP observation, they induce  \,$K\to\pi X$\,
at rates larger than the existing bounds;
\vspace*{-1ex}
\item if they are dominated by loop diagrams involving up-type quarks and $W$ bosons, they
have a~$CP$ phase that is too large;
\vspace*{-1ex}
\item if they are dominated by loop diagrams involving up-type quarks and $W$ bosons, their
scaling to the $B$ system is incompatible with the nonobservation of  \,$B\to X_sX$.
\vspace*{-1ex}
\end{list}
We have found in this paper that (a) can be solved in some cases by the addition of the effects
of four-quark operators. We have suggested that more general models may be constructed to solve
(b) and (c). To show that this is possible, we have constructed a specific example in
Ref.~\cite{usletter}.

Disregarding existing bounds from kaon and $B$-meson decays, we have shown that many light
Higgs bosons have couplings of the right size to explain the HyperCP observation.
We think this is sufficiently intriguing to warrant a revisiting of the kaon and $B$ decay
results.
In particular, the $B$ factories are still operational and could reanalyze the very low
$m_{\mu\mu}^{}$ invariant-mass region in their measurements of  \,$B\to X_s\mu^+\mu^-$\, modes.
The NA48 experiment might also be able to revisit the kaon modes.

\begin{acknowledgments}

The work of X.G.H. was supported in part by NSC and NCTS. The work of G.V. was supported
in part by DOE under contract number DE-FG02-01ER41155.
We thank Laurence Littenberg and Rainer Wanke for useful discussions on the kaon bounds and
Soeren Prell for useful discussions on the $B$ bounds.

\end{acknowledgments}

\end{document}